\documentclass[12pt]{article} %,a4paper

\usepackage{graphicx}
\usepackage{amsmath}
\usepackage{amssymb}
\usepackage{hyperref}
\usepackage[height=8.8in,width=6.45in]{geometry}
\usepackage{tikz}
\usepackage{cite}
\usepackage{amscd}
\usepackage{setspace}
\usepackage{bbm}
\usepackage{dsfont}
\usepackage{cancel}
\usepackage{relsize}
\usepackage[utf8]{inputenc}
\usepackage{xcolor}
\usepackage[normalem]{ulem}
\numberwithin{equation}{section}

\newcommand{\cO}{\mathcal{O}}

\newcommand{\cQ}{{\cal Q}}
\newcommand{\cS}{\mathcal{S}}

\newcommand{\QK}{\mathrm{QK}}
\newcommand{\QH}{\mathrm{QH}}
\newcommand{\GL}{\mathrm{GL}}
\newcommand{\Gr}{\mathrm{Gr}}

\newtheorem{conj}{Conjecture}
\newtheorem{example}{Example}
\newtheorem{example*}{Example}
\newtheorem{example**}{Example}
\newtheorem{example***}{Example}

\begin{document}
\begin{titlepage}

\begin{flushright}

\end{flushright}

\vskip 3cm

\begin{center}
{\Large \bf
 A correspondence between the quantum K theory and quantum cohomology of Grassmannians}

\vskip 2.0cm

Wei Gu$^{1,2}$, Jirui Guo$^3$, Leonardo Mihalcea$^{4}$, Yaoxiong Wen$^5$, Xiaohan Yan$^6$ \\

\bigskip
\bigskip

%\begin{tabular}{cc}
%$^1$ Bethe Center for Theoretical Physics, Universit{\"a}t Bonn,  
% D-53115 Bonn, Germany \\
%$^2$ Max-Planck-Institut f{\"u}r Mathematik, Vivatsgasse 7, D-53111 Bonn, Germany\\
% $^3$ School of Mathematical Sciences and Institute for Advanced Study\\ 
% Tongji University, Shanghai, 200092, China\\
% $^4$ Department of Mathematics, Virginia Tech, MC 0123 225 Stanger Street\\
% Blacksburg, VA, 24061, USA\\
%$^5$ Korea Institute for Advanced Study,  85 Hoegiro, Dongdaemun-gu, Seoul, 02455\\ \ \ Republic of Korea.\\
% $^6$ Sorbonne Universit{\'e} and Universit{\'e} Paris Cit{\'e} CNRS, IMJ-PRG F-75005, Paris, France\\
%\end{tabular}

\begin{tabular}{ll}
$^1$ Max-Planck-Institut f\"ur Mathematik, Vivatsgasse 7,
D-53111 Bonn, Germany\\
$^2$ Bethe Center for Theoretical Physics, Universit\"{a}t Bonn,  D-53115 Bonn, Germany \\
 $^3$ School of Mathematical Sciences and Institute for Advanced Study, Tongji University \\ \ \ Shanghai, 200092, China\\
 $^4$ Department of Mathematics, MC 0123 225 Stanger Street, Virginia Tech, Blacksburg\\ \ \ VA, 24061, USA\\
$^5$ Korea Institute for Advanced Study,  85 Hoegiro, Dongdaemun-gu, Seoul, 02455\\ \ \ Republic of Korea\\
 $^6$ Sorbonne Université and Université Paris Cité CNRS, IMJ-PRG F-75005 Paris, France\\
\end{tabular}

\vskip 1cm

\textbf{Abstract}

\end{center}

  We utilize physics arguments, and the nonabelian/abelian correspondence, to relate the Givental and Lee's quantum K theory ring of Grassmannians to a twisted variant of the quantum cohomology ring. Furthermore, the quantum K pairing is related to correlators arising from supersymmetric localization. We state some mathematical conjectures, which we illustrate in several examples.

\medskip
\noindent
\bigskip
\bigskip
\bigskip
\begin{flushright}
  \textit{Dedicated to the memory of Bumsig Kim, whose work continues to inspire us.}
\end{flushright}

\bigskip
\vfill
\end{titlepage}

\setcounter{tocdepth}{2}
\tableofcontents

\section{Introduction}\label{Int}

{Let $X=V/\!/\mathds{G}$ be a GIT quotient defined by a vector space $V$, and a reductive group $\mathds{G}$. One of the recent tools to study Gromov-Witten theory of GIT quotients is the 
\emph{nonabelian/abelian correspondence}. Roughly, it states that the Gromov-Witten theory of the `nonabelian space' $V/\!/\mathds{G}$ is related to that of the `abelian space' $V/\!/\mathds{T}$, where $\mathds{T}$ is a maximal torus in $\mathds{G}$. The classical version goes back to Ellingsrud and Str{\o}mme \cite{ES89}, and Martin \cite{Martin:2000}, and the quantum formulation is due to Bertram, Ciocan-Fontanine, and Kim \cite{BCK05,BCK08} and Ciocan-Fontanine, Kim, and Sabbah \cite{CKS08}. In this paper we consider the GIT quotient $X$ to be the Grassmannian $\mathrm{Gr}(k;N)$. Continuing earlier work on quantum K theory of partial flag manifolds \cite{Gu:2023tcv,Gu:2023math}, and motivated by the formalism of the nonabelian/abelian correspondence, in this paper we investigate a relation between the quantum K theory 
ring of $\mathrm{Gr}(k;N)$, defined by Givental and Lee \cite{Givental:2000, Lee:2001mb}, and a certain twisted
version of the quantum cohomology ring of $\mathrm{Gr}(k;N)$. Our aim is to shed light on the role played by the nonabelian/abelian correspondence in relating the Frobenius structures - especially the pairings - of quantum K theory and (twisted) quantum cohomology.}

 The nonabelian/abelian correspondence also appears naturally in physics, in the context of supersymmetric gauge theories. Indeed, the semiclassical vacuum configuration in nonabelian supersymmetric gauge theories suggests that the gauge group may be Higgsed to a semi-product of its maximal torus and Weyl group \cite{Witten:1993xi, Seiberg:1994rs, Hori:2000kt, Hori:2006dk, Halverson:2013eua}, which we call the associated abelian-like theories. Therefore, the nonabelian/abelian correspondence can be rephrased as the physical statement that a nonabelian gauge theory and its associated abelian-like theory share the same vacuum structure and BPS spectrum. For 2d ${\cal N}=(2,2)$ nonabelian gauge theory, a physical discussion is provided in \cite{Gu:2018fpm}. The physical incarnation
 %re-incarnation 
 above suggests that one may use various results in supersymmetric gauge theories 
 \cite{Benini:2015noa, Closset:2015rna, Closset:2016arn, Pestun:2016zxk, Gu:2020ana} 
 to explore new conjectural connections in mathematics.  
 {We mention that Kapustin and Willet \cite{Kapustin:2013hpk} have found 
 that a certain variant of the quantum K theory for the Grassmannian 
 corresponds to the quantum cohomology of the same target,
 by observing that they are both isomorphisms to the associated 
 Verlinde algebra. However, it turns out that this variant of the quantum K ring is not the same
 as that defined by Givental and Lee.}

{We now turn to more precise versions of our results. In this paper we propose a relation between Givental-Lee's quantum K theory and the quantum cohomology of Grassmannians. Since quantum K theory and quantum cohomology can be understood as topological sectors of, respectively, the 3d ${\cal N}=2$ Chern-Simons-matter theory on spacetime $\Sigma\times S^1$ and the 2d ${\cal N}=(2,2)$ gauge theory, we utilize the exact results of correlation functions of supersymmetric gauge theories to prove the relation.  

Our first result is that the correlators of the quantum K theory can be computed via certain twisted correlation functions in a \textit{twisted quantum cohomology} (tQH) ring:
%{\color{red}
%\begin{equation}\label{}\nonumber
%  \left\langle \mathcal{O} (z,t) \right\rangle^{{\rm QK}}= \left\langle\mathcal{O} (L\sigma_{a}, t)\star \left(\det(1-L\sigma_{a})\right)^{-k} \right\rangle^{{\rm tQH}},
%\end{equation}
%} 

\begin{equation}\label{}\nonumber
  \left\langle \mathcal{O} (x | t) \right\rangle^{{\rm QK}}= \left\langle\mathcal{O} \bigl(1-L\sigma | 1-L m\bigr)\star \left(\det(1-L\sigma)\right)^{-k} \right\rangle^{{\rm tQH}} \/.
\end{equation}
%\begin{equation}\label{}\nonumber
 % \left\langle\prod_{\lambda}G_{\lambda}(z,t) \right\rangle^{{\rm QK}}= \left\langle\prod_{\lambda}G_{\lambda}%(L\sigma_{a}, t)\star \left(\det(1-L\sigma_{a})\right)^{-k} \right\rangle^{{\rm tQH}},
%\end{equation}
(We expect the insertion factor on the right-hand side to be 
$\det(1-L\sigma_{a})^{k(g-1)}$ for the higher genus cases, where $L$ is the perimeter of $S^1$.) 
%The insertions on the left-hand side are chosen in $\mathrm{QK}_{G}(\mathrm{Gr}(k;N))$ the (double) Grothendieck polynomials defined in (\textbf{Add-refs}),
{Here, the observable $\mathcal{O} (x|t)$ is a polynomial in the variables $x=(x_1, \ldots, x_k)$,
and the equivariant parameters $t=(t_1, \ldots, t_N)$ of the natural 
$(\mathbb{C}^\times)^N$-action on $\mathrm{Gr}(k;N)$.}
The variables $\sigma=(\sigma_a)$ are the Chern roots of $\mathcal{S}$, the tautological rank $k$ subbundle on $\mathrm{Gr}(k;N)$, i.e., they satisfy $x_a = e^{-L \sigma_a}$, with $L=2 \pi R$ a homogenization parameter.\begin{footnote}{ We identify $x_a$ with $e^{-2\pi R \sigma_a}$ by taking the Chern character. We will see in section \ref{GLSMACS} that the (small) quantum K theory can be constructed from the 3d ${\cal N}$=2 Chern-Simons theory on spacetime $\Sigma\times S^1$, where $R$ is the radius of $S^{1}$. The perimeter $L:=2 \pi R$ is often chosen to be equal to $1$ in the math literature. However, since it has the opposite degree as the one of $\sigma_a$, we keep this parameter explicitly such that $x_{a}$ is a well-defined zero degree quantity.}\end{footnote}
Similarly, the masses $m=(m_i)$ satisfy $t_i = e^{-L m_i}$. We refer to \eqref{E:tQK-cor} and \eqref{E:maintQH} for precise definitions. 

From the definition it follows that the twisted correlator is a certain (Jeffrey-Kirwan) residue, obtained from supersymmetric localization 
(cf.~\eqref{E:tQK-cor} and \eqref{E:etQK-cor} below), and 
implying it is a `top form' $\oint_{Gr(k;N)} \cO(x|t)$, which picks up 
the coefficient of the point in a certain {\em cohomological} expansion 
into Schubert classes of a polynomial representative of $\cO(x|t)$; 
see \eqref{E:topform} and \eqref{E:eqtopform} below. 
The proof of this uses the nonabelian/abelian correspondence in cohomology, and the tQH correlator is conjecturally related to a twisted quantum cohomology ring, discussed below. 

However, it turns out that the tQH correlator may also be calculated via a holomorphic Euler characteristic of a quantum K product, extended by linearity with respect to the quantum parameter $q_{3d}$. The precise statements are given in Conjectures \ref{conj:pairing} and \ref{conj:pairing2} below, and are accompanied by several worked out examples. In this context we have an equality
\begin{equation*} \left\langle \cO(x|t) 
\right\rangle^{\mathrm{tQH}} = \langle \cO(x|t) \star(\det\mathcal{S})^{\star k} \rangle^{\mathrm{QK}} \/,\end{equation*}
which in turn gives a new formula to calculate the QK pairing $\langle \cdot , \cdot \rangle$ in the quantum K theory. If $a,b \in \mathrm{K}_{\mathds{T}}(\Gr(k;N))$, then 
\[ \langle a, b \rangle = \oint_{Gr(k;N)} \frac{a \star b}{(\det \mathcal{S})^{\star k}}= \frac{1 }{\Bigl((1-q_{3d}) \cdot \prod_{i=1}^N t_i \Bigr)^k}  \oint_{Gr(k;N)} a \star b \star \det (\mathcal{Q})^{\star k} \/, \]
where $\mathcal{Q}$ is the quotient bundle on $\Gr(k;N)$. 
The relation $\det \mathcal{S} \star \det \mathcal{Q} = (1-q_{3d}) \prod_{i=1}^N t_i$ 
in the quantum K theory ring $\mathrm{QK}_{\mathds{T}}(\Gr(k;n))$ (see \cite{Gu:2022yvj})
allowed us to remove the negative power of the determinant in the denominator. 
A different formula for the quantum K pairing $\langle a, b \rangle$ has been found in 
\cite{Buch2020}. See also \cite{Jockers:2018sfl, Jockers:2019lwe, Ueda:2019qhg,Closset:2023bdr} for the recent discussions via physics.

 At the level of rings, we start with the K-theoretic version of the vacuum (or Bethe Ansatz) equations, which are known from \cite{Gorbounov:2014,Gu:2022yvj} that give the relations in the quantum K ring: 
\begin{equation}\label{}\nonumber
 \sigma_{a}^{N}=(-1)^{k-1}q\frac{(1-L\sigma_{a})^{k}}{\prod^{k}_{b=1}(1-L\sigma_{b})}.
\end{equation}
Compared to cohomology, their right hand side has a twisting factor which comes from contributions from Chern-Simons terms, and which has no direct origin in the two-dimensional gauge theory. To recover the effect of this factor in quantum cohomology, motivates us to define a quotient ring of the ordinary equivariant quantum cohomology.
More specifically, the global symmetry of the Grassmannian ${\rm Gr}(k;N)$ is denoted by $G(=SU(N))$. Our new observation is that there is an isomorphism of rings
\[ \mathrm{QK}_{G}(\mathrm{Gr}(k;N)) \simeq \widetilde{\mathrm{QH}}^{\ast}_{G}( \mathrm{Gr}(k;N)) \]
which transforms the equivariant parameters in a non-trivial, but explicit, way. See \eqref{DM} in the non-equivariant case; the equivariant case is given in \eqref{DM2}. The ring on the right-hand side is a quotient of the $G$-equivariant quantum cohomology ring $\mathrm{QH}^\ast_G(\mathrm{Gr}(k;N))$ by an ideal generated by explicit polynomials; see sections \ref{NEC} and \ref{NC}. We also call this ring the twisted quantum cohomology (tQH) ring.  To build up the dictionary between the two rings, we choose in $\mathrm{QK}_{G}(\mathrm{Gr}(k;N))$ the classes corresponding to the $G$-equivariant `Grothendieck bundles'
%{\color{blue} generators of K-rings}
\footnote{Although they are useful for finding the dictionary between the quantum K theory and the quantum cohomology, their geometric meanings are still unclear at the moment.\cite{Gu:2022yvj} defined by a symmetric variant of the (double) Grothendieck polynomials:
%represented by

\begin{equation}\label{}\nonumber
G'_{j}(1-x,1-t)=\sum_{j_1+j_2=j}(-1)^{j_1}e_{j_1}(1-t)G_{j_2}(1-x) \/;
\end{equation}

where $e_{j_1}(1-t)\in {\rm K}_G(pt)$ are the 
elementary symmetric polynomials in $1-t_1,\ldots, 1-t_N$ and $G_{j_2}(1-x)$ are the 
Grothendieck polynomials in $1-x_1, \ldots, 1-x_k$. The more general polynomial 
$G'_{j}(1-x,1-t)$ is a deformation of the factorial Grothendieck polynomial from \cite{McNamara:2005}, which is symmetric in both $x$ and $t$ variables. 
On the cohomological side, denote by $\sigma = (\sigma_a)_{a=1}^k$ 
the Chern roots of the tautological bundle and by $m = (m_i)_{i=1}^N$ 
the (cohomological) equivariant parameters.

Then, our dictionary (see Section \ref{QKQHC} for more details) connects
\begin{equation}\label{E:change-of-vars}%\nonumber
  G'_{j}(1-x,1-t) \leftrightarrow L^{j}{\cal S}_{j}(\sigma|m) \/,
\end{equation}
where the perimeter of $S^{1}$ is denoted by $L=2\pi R$, and 
${\cal S}_{j}(\sigma_{a}| m) \in {\rm QH}^{\ast}_{{G}} (\mathrm{Gr}(k;N))$
are represented by the $G$-equivariant ``factorial Schur functions".\footnote{For $\mathrm{Gr}(k;N)$, the target space symmetry $G$ is the $SU(N)$ group, so its maximal torus $T$ is $U(1)^{N-1}$. A notion of the so-called $T$-equivariant factorial Schur functions factor has appeared e.g. in \cite{Mihalcea:2008}, see also references therein. Here, our basis can be regarded as a particular sector of the $T$-equivariant factorial Schur functions by imposing the $S_N$-invariant constraint on the basis.} Note that the relation between the K-theoretic and cohomological equivariant parameters is less direct. Even when we take $t_i=1\ (\forall i)$ and consider the non-equivariant quantum K theory, the equivariant parameters in quantum cohomology still persist (cf. \eqref{DM}):
\begin{equation}\label{}\nonumber
  G_{j}(z) \leftrightarrow L^{j}{\cal S}_{j}(\sigma|m)\/.
\end{equation}

In fact, the dictionary between the two theories is discovered by matching the quantum ring relations, and the cohomological equivariant parameters are necessary to make it possible.}

Finally, one should keep in mind that our dictionary is only defined on the vacua of the associated quantum field theories, so it makes sense to physically build a map between the expectation values of $\sigma$ fields with the mass parameters. From the field-theoretic perspective, it means we have to fine-tune the parameters in quantum cohomology to identify the topological data of the two theories. {Another well-known point of view in quantum field theory is that these twisted masses can be obtained by first weakly gauging the global symmetry group $G$ acting on the theory, coupling the matter fields to a vector multiplet for $G$, and then one could add more constraints into the Lagrangian of this theory such that the dictionary can be gained from the vacuum equations of this ``primary" quantum field theory.
The correlators of the associated twisted quantum field theory can also be understood as the correlators of this ``primary" quantum field theory, although its study will appear in future work.

This paper is organized as follows: In section \ref{SGTF}, we review the topological aspects of topological supersymmetric gauge theory with four supercharges. More concretely, the quantum cohomology can be understood as the topological sector of a 2d ${\cal N}=(2,2)$ gauge theory; the quantum K theory can be formalized from the topological correlators of the Wilson loops of 3d ${\cal N}=2$ Chern-Simons-matter theory on space-time $S^{1}\times\Sigma$. In section \ref{QKQHC}, we present our new observation about the connection between the quantum K theory and quantum cohomology.

\section{Mathematical aspects of supersymmetric gauge theories with four supercharges}\label{SGTF}

This section is devoted to describing some mathematical background of supersymmetric gauge theories. Readers familiar with these facts can safely skip this section.

\subsection{Quantum cohomology}\label{QH}
For a
%given
K\"{a}hler manifold $X$, the quantum cohomology ring is a deformation
%an extension
of the ordinary cohomology ring. In this paper, we focus mostly on the case where $X$ is a complex Grassmannian. Deformations of $X$ are parametrized by the ``complexified K\"{a}hler moduli space" which we denote by ${\cal M}_{K}$. Moreover, we denote the cohomology ring of $X$ by $\mathrm{H}^{\ast}(X)=\oplus_{n\in\mathbb{N}}\mathrm{H}^{2n}(X)$. The associative cup product defines a nondegenerate bilinear pairing $\langle \cdot , \cdot \rangle$ on $\mathrm{H}^{\ast}(X)$:
\begin{equation}\label{PAIR}
  g_{ij}=\langle T_{i},T_{j}\rangle:=\int_{X}T_{i}\cup T_{j}, \quad\quad T_{i}\in \mathrm{H}^{\ast}(X).
\end{equation}
The cohomology ring $\mathrm{H}^\ast(X)$ and the pairing provide a structure
of a Frobenius algebra.
If $T_{i}, T_{j}, T_{k}$ are represented by algebraic cycles, the
\textit{structure constants}
$\langle T_{i}, T_{j},T_{k}\rangle:=\langle T_{i}\cup T_{j},T_{k}\rangle=\langle T_{i}, T_{j}\cup T_{k}\rangle$
of the cup product count
the number of intersections of $T_{i}$, $T_{j}$, $T_{k}$ translated in generic position.
These structure constants can be extended to those of the so-called \textit{(small) quantum cup product} by
\begin{equation}\label{QHP}
  C_{ijk} = \left\langle T_{i}\star T_{j}, T_{k}\right\rangle := \sum_{d\geq 0}q^{d}\left( T_{i}, T_{j},T_{k}\right)_{d},
\end{equation}
where $q$ is the ``instanton parameter":
\begin{equation}\label{IP}
  q: {\cal M}_{K}\times \mathrm{H}_{2}(X)\rightarrow\mathbb{C}\quad\quad \left(\omega,\beta\right)\mapsto q^{d}=e^{2i\pi\int_{\beta}\omega}.
\end{equation}
Denote by $\overline{\cal M}_{3,d}$ the moduli space of genus-zero stable maps with three marked points to $X$ compactifying degree $d$ maps $\mathbb{CP}^{1}\rightarrow X$. Then, the Gromov-Witten (GW) invariant with three insertions $\left(T_{i}, T_{j},T_{k}\right)_{d}$ , or the ``three-point correlation function'', is defined as
\begin{equation}\label{DefGW}
\left(T_{i}, T_{j},T_{k}\right)_{d} := \int_{[\overline{\cal M}_{d}]^{\text{vir}}} ev_1^{\ast}T_i \cup ev_2^{\ast}T_j \cup ev_3^{\ast}T_k,
\end{equation}
where $ev_l\ (l=1,2,3)$ are the evaluation maps and $[\overline{\cal M}_{d}]^{\text{vir}}$ is the virtual fundamental cycle (VFC). Note that the right-hand side of Eq.(\ref{QHP}) is only non-vanishing when the degree of integrand matches that of the VFC, giving the ``grading'' in quantum cohomology. From Eq.(\ref{PAIR}) and Eq.(\ref{QHP}), the quantum product takes the form
\begin{equation}\label{SCC}
    T_{i}\star T_{j}:=C^{k}_{ij}T_{k}=C_{ijk}g^{kl}T_{l}.
\end{equation}
One can see that the above definition for small quantum cohomology needs only three-point correlation functions. In contrast, in big quantum cohomology, one needs all higher-point correlation functions to define the structure constants.

Relations of the quantum cohomology ring are deformations of the classical relations.
For the example of projective space $\mathbb{CP}^{N-1}$, the classical ring relation is $H^{N}=0$, where $H \in \mathrm{H}^{2}(X)$ is the generator of the cohomology Poincaré-dual to the projective hyperplane class. In quantum cohomology, this relation
%it
is deformed by the ``instanton number'' into the quantum ring relation $H^{N}=q$.

The quantum cohomology is ``rigid'' in the sense that it can only be
further deformed by the equivariant parameters associated with the
global symmetry of the target $X$.
We denote the torus-equivariant quantum cohomology by
$\mathrm{QH}^{\star}_{T}(X)$, for $T$ the maximal
torus of the global symmetry $G$ of $X$. Furthermore, denote the
Weyl-symmetry of group $G$ as $W$,
then we have
$\mathrm{QH}^{\ast}_{G}(X)=(\mathrm{QH}^\ast_T(X))^W$, so
$\mathrm{QH}^{\ast}_{G}(X)$ is the sub-ring of $\mathrm{QH}^\ast_T(X)$
consisting of elements invariant under the Weyl-symmetry.
(This follows from the more familiar fact that $\mathrm{H}^\ast_G(X) = \mathrm{H}^\ast_T(X)^W$; see e.g., \cite{brion:eqint}.)
In most cases of interest (including the complex Grassmannian), making the torus equivariant parameters equal to $0$ recovers the ordinary cohomology.
\begin{example}\label{Ex:1} Let $X = {\rm Gr}(k;N)$ be the Grassmannian parametrizing linear subspaces of dimension $k$ in $\mathbb{C}^{N}$. We have a tautological short exact sequence $0\rightarrow {\cal S}\rightarrow\mathbb{C}^{N}\rightarrow {\cal Q}\rightarrow 0$, where the rank of the ${\cal S}$ bundle is $k$. The non-equivariant quantum cohomology of Grassmannian can be represented \cite{Witten:1993xi} as
\begin{equation}\label{}\nonumber
  \mathrm{QH}^{\ast}({\rm Gr}(k;N))=\frac{\mathbb{Z}[q]\left[e_{1}(\sigma),\ldots,e_{k}(\sigma);e_{1}(\widetilde{\sigma}),\ldots,e_{n-k}(\widetilde{\sigma})\right]}{\left\langle\left(\sum^{k}_{i=0}e_{i}(\sigma)\right)\left(\sum^{N-k}_{i=0}e_{i}\widetilde{\sigma}\right)=1+(-1)^{k}q\right\rangle}.
\end{equation}
{where $\sigma$ and $\widetilde{\sigma}$ are the Chern roots of $\mathcal{S}$ and $\mathcal{Q}$ respectively.}
 In this setup, the equivariant quantum cohomology can be
 %easily
 represented \cite{Kim:1995, Mihalcea:2008, Gu:2022yvj}
 \begin{equation}\label{}\nonumber
  \mathrm{QH}^{\ast}_{{ T}}({\rm Gr}(k;N))=\frac{{\mathbb{Z}[m_1, \ldots, m_N]} \left[e_{1}(\sigma),\ldots, e_{k}(\sigma);e_{1}(\widetilde{\sigma}),\ldots,e_{n-k}(\widetilde{\sigma})\right]}{\left\langle\left(\sum^{k}_{i=0}e_{i}(\sigma)\right)\left(\sum^{N-k}_{i=0}e_{i}(\widetilde{\sigma})\right)=\sum^{N}_{i=1}e_{i}(m)+(-1)^{k}q\right\rangle},
\end{equation}
where $m = (m_i)_{i=1}^N$ are the equivariant parameters, or the so-called twisted masses in physics.
\end{example}
The special mathematical phenomenon of nonabelian/abelian correspondence says that the (small) quantum cohomology of ${\rm Gr}(k;N)$ can be extracted from that of its associated toric variety $\mathbb{CP}^{N-1}\times\cdots\times \mathbb{CP}^{N-1}$ ($k$ factors)\cite{Martin:2000, BCK05, Hori:2000kt, Gu:2020ana}. We end this section by remarking that the small quantum cohomology can be understood from the topological quantum gauge theory (to be reviewed in section \ref{GLSMACS}), where the nonabelian/abelian correspondence is \emph{manifest} in gauge theory.

\subsection{Quantum K theory}\label{QKT}
We first recall that the classical K theory of a complex variety $X$ is generated by complex vector bundles. The equivariant K ring $\mathrm{K}_{T}(X)$ of the manifold $X$ is generated additively by the symbols $[F]$, where $F\rightarrow X$ is any $T$-equivariant vector bundle, modulo the relations $[F]=[F_{1}]+[F_{2}]$ for any short exact sequence of vector bundles $0\rightarrow F_{1}\rightarrow F\rightarrow F_{2}\rightarrow0$. The additive structure on $\mathrm{K}_{T}(X)$ is given by direct sum, and the multiplication is given by the tensor product of vector bundles. For simplicity of notation, we recall the Hirzebruch $\lambda_{y}$ class which is useful in various settings
\begin{equation}\label{KR}
    \lambda_{y}(F):=1+y[F]+y^{2}[\wedge^{2}F]+\ldots+y^{k}[\wedge^{k}F]\in {\rm K}_{T}(X)[y],
\end{equation}
where $k$ is the rank of $F$. It can be thought of as a K theoretic analogue of the (cohomological) Chern polynomial
\begin{equation}\label{CP}
    c_{y}(F):=1+yc_{1}(F)+\ldots+y^{k}c_{k}(F)
\end{equation}
of the bundle $F$.~The ring $\mathrm{K}_{T}(X)$ is naturally an algebra over
$\mathrm{K}_{T}(pt)={\rm Rep}(T)$,
the representation ring of $T$, which is the Laurent polynomial ring
$\mathrm{K}_{T}(pt)=\mathbb{Z}[t^{\pm}_{1},\ldots,t^{\pm}_{N}]$
%since $T$ is a torus,
with $t_{i}$ being characters corresponding to an integral basis of the Lie algebra of $T$.

Since $X$ is proper, the K-theoretic push-forward to a point is well-defined,
and it equals the (holomorphic) Euler characteristic
\begin{equation}\label{EC}
\chi(X,F):= \sum_{i}(-1)^{i} H^{i}(X,F) \in \mathrm{Rep}(T).
\end{equation}
Moreover, the ``classical" pairing is defined as
\begin{equation}\label{PR}
  \left\langle -,-\right\rangle_0: {\rm K}_{T}(X)\otimes{\rm K}_{T}(X)\rightarrow {\rm K}_{T}(pt);\quad\quad \langle[F],[E]\rangle_0 := \chi\left(X, F\otimes E\right).
\end{equation}

The Givental-Lee's quantum K theory is defined in \cite{Givental:2000, Lee:2001mb}. Assume $\{{\cal O}_\lambda\}$
is a basis of ${\rm K}_{T} \left({\rm Gr}(k;N)\right)$ over $\mathrm{K}_{T}(pt)$, where $\lambda$ varies over partitions in the $k\times (N-k)$ rectangle. Unlike quantum cohomology, there is no ``grading" induced from dimension arguments on the moduli space in (small) quantum K theory. Moreover, the natural pairings on quantum K-ring, defined a priori in terms of the K-theoretic two point correlation functions, admit now nontrivial quantum corrections compared to the classical pairing above:
\begin{equation}\label{QP}
 \langle {\cal O}_{i}, {\cal O}_{j}\rangle:=\sum_{d\geq 0}q_{3d}^{d}\left({\cal O}_{i}, {\cal O}_{j}\right)_{d}\quad\in {\rm K}_{T}\left(pt\right)[\![q_{3d}]\!].
\end{equation}
The pairing is extended to ${\rm K}_{T} \left({\rm Gr}(k;N)\right)[\![q_{3d}]\!]$ by ${\rm K}_{T}\left(pt\right)[\![q_{3d}]\!]$-linearity. On the right-hand side are the K-theoretic two-point correlation functions (K-theoretic GW invariants)
%like
$\left({\cal O}_\lambda, {\cal O}_\mu\right)_{d}$, defined as holomorphic Euler characteristics on the moduli spaces $\overline{{\cal M}}_{d}$ \cite{Givental:2000, Lee:2001mb}. More precisely,
%\[
%\left({\cal O}_\lambda, {\cal O}_\mu\right)_{d} :=
%\chi(\overline{{\cal M}}_{d}, \mathcal{O}^{\text{vir}}\otimes ev_1^*\mathcal{O}_\lambda\otimes ev_2^*\mathcal{O}_\mu),
%\]

\[
    \left({\cal O}_\lambda, {\cal O}_\mu\right)_{d} :=\begin{cases}
    	\chi({\rm Gr}(k;N), {\cal O}_\lambda \otimes {\cal O}_\mu), \quad d=0,\\
    	\chi(\overline{{\cal M}}_{2, d}, \mathcal{O}^{\text{vir}}\otimes ev_1^*\mathcal{O}_\lambda\otimes ev_2^*\mathcal{O}_\mu), \quad d \geq 1,
    \end{cases}
\]
Since we consider $\overline{{\cal M}}_{2, d}$, we need to be careful when $d=0$.
where $\mathcal{O}^{\text{vir}}$ is the virtual structure sheaf on $\overline{{\cal M}}_{d}$ and $ev_l\ (l=1,2)$ are the evaluation maps. For Grassmannians, the virtual sheaf is equal to the fundamental class of the moduli space.
\begin{example} Consider the projective space $\mathbb{CP}^{2}$. For simplicity, we focus on the computation without $T$-action. For the Schubert basis ${\cal O}_{0}=[{\cal O}_{\mathbb{CP}^{2}}]$, ${\cal O}_{1}=[{\cal O}_{line}]$, ${\cal O}_{2}=[{\cal O}_{pt}]$, the classical K pairing is given by
\begin{equation}\nonumber
\{ \left({\cal O}_{i},{\cal O}_{j}\right)_{0} \}_{i,j=0}^2 =\left(
  \begin{array}{ccc}
    1 & 1 & 1 \\
    1 & 1 & 0 \\
    1 & 0 & 0 \\
  \end{array}
\right).
\end{equation}
For any $i,j\geq 0$ and $d>0$, $\left({\cal O}_{i}, {\cal O}_{j} \right)_{d}=1$. From
\cite{Buch:2008} or \cite{Buch2020} one may calculate
%Hence,
the quantum K pairing
%is given
by
\begin{equation}\nonumber
   \langle {\cal O}_{i}, {\cal O}_{j}\rangle = \left( {\cal O}_{i}, {\cal O}_{j}\right)_{0}+\frac{q_{3d}}{1-q_{3d}}.
\end{equation}
\end{example}
Now we are ready to define the quantum K product $\star$ on $\mathrm{QK}_{T}(X)$ via the (K-theoretic) three-point functions, defined similarly to the two-point functions above. We require by definition that ${\cal O}_{i}\star {\cal O}_{j}$ satisfies
\begin{equation}\label{QKP}
    \langle{\cal O}_{i} \star {\cal O}_{j}, {\cal O}_{k}\rangle=\sum_{d\geq 0}q_{3d}^{d}\left({\cal O}_{i}, {\cal O}_{j}, {\cal O}_{k}\right)_{d},
\end{equation}
for any ${\cal O}_{i}, {\cal O}_{j}, {\cal O}_{k}\in\mathrm{K}_{T}(\mathrm{Gr}(k;N))$. See \cite{Buch:2008} for more setup and information. It was proved in \cite{Gu:2022yvj} that the quantum K theoretic ring relation of Grassmannian can be expressed in terms of the $\lambda_{y}$ classes as follows:
\begin{equation}\label{QKRL}
  \lambda_{y}({\cal S})\star \lambda_{y}({\cal Q})=\lambda_{y}(\mathbb{C}^{N})-\frac{q_{3d}}{1-q_{3d}}y^{N-k}\left(\lambda_{y}\left({\cal S} \right)-1\right)\star\det {\cal Q}.
\end{equation}
One advantage of this expression is that the equivariant parameters of the $T$-action will be stored solely in the factor $\lambda_{y}(\mathbb{C}^{N})$.\\

\noindent{\textbf{Some comments}}. One may expect that the quantum K theory is as rigid as the quantum cohomology, in the sense that one cannot deform it beyond the equivariant deformation. However, due to the lack of grading in K theory, we may perform nontrivial modification on the quantum K theory by tensoring the virtual structure sheaves with (arbitrary powers of) certain determinant line bundles on the moduli space of stable maps. This leads to the notion of quantum K theory with level structures. Roughly,
%We formally denote the determinant bundle by ``det''.
the level $l$ K-theoretic GW invariants roughly take the form
\begin{equation}\label{LEC}
\chi\left(\overline{{\cal M}}_{n,d}, \otimes_{i=1}^n ev_i^* (F_i) \right)^{l} := \chi(\overline{{\cal M}}_{n,d}, \mathcal{O}^{\text{vir}}\otimes {\det}^l \otimes \otimes_{i=1}^n ev_i^* (F_i)) \/,
\end{equation}
where $\det$ denotes the determinant of an appropriate bundle.
The pairing and the quantum product of the quantum K ring with level structures are modified in the same pattern. The reader is referred to \cite{Ruan2018} for a more complete mathematical formulation. Physicists predict \cite{Kapustin:2013hpk, Witten:1993xi} that with a particular choice of level structure, a ``grading" emerges on the quantum K ring so modified, and that the ring becomes isomorphic to the quantum cohomology. We will see in section \ref{GLSMACS} that the {QK ring} corresponds to a special choice of the gauge-Chern-Simons level. Finally, we would like to point out that several physical concepts still lack parallel definition in mathematics. For example, the level structure would affect the dimension of the state space in physics, but its mathematical interpretation is still missing.\\

\noindent{\textbf{Nonabelian/abelian correspondence}}. On the other hand, the nonabelian/abelian correspondence in quantum K theory, unlike that in quantum cohomology, is still not fully understood. For example, as we will see in section \ref{GLSMACS}, by arguments of quantum field theory, the quantum K ring relations of Grassmannian may not be obtained directly from those of the associated abelian space $(\mathbb{CP}^{N-1})^{\times k}$ in the same way as those in quantum cohomology. Instead, it is the quantum K ring with a certain level structure, which corresponds to the Verlinde algebra, that may serve as the suitable candidate on the abelian side. Indeed, the K-theoretic nonabelian/abelian correspondence for Grassmannians is still an open question.\begin{footnote}{ While this paper was in final stages of preparation, the paper \cite{IHK:twisted} appeared, which shows how the relations in the QK theory of Grassmannians arise as relations of a certain twisted QK theory
of the abelianization.}\end{footnote}
In this paper, we attack this problem through a physics argument, read off from quantum field theory, by identifying Givental-Lee's quantum K theory with a specialized equivariant quantum cohomology. We leave the rigorous mathematical treatment to {future work}.
%\cite{Gu24}.

In the section below, we will review some basics of quantum field theory that are useful for presenting our observation.

\subsection{GLSMs and 3d ${\cal N} = 2$ Chern-Simons-matter theories} \label{GLSMACS}
The initial understanding of quantum cohomology in quantum field theory can be formalized in terms of the A-twisted nonlinear sigma model (NLSM) \cite{Witten:1988xj} with the Lagrangian
\begin{equation}\label{NLSM}
  S=\int d^{2}xd^{4}\theta {\cal K}(\Phi,{\bar \Phi}),
\end{equation}
where ${\cal K}$ is the K\"{a}hler potential of the target $X$. Each $\Phi$ is a chiral superfield with the lowest component being the coordinates of the target space $X$. It is a 2d ${\cal N}$=(2,2) supersymmetric theory. Following Witten, one can perform the so-called topological A-twist, which generates a scalar supercharge $Q_{A}$. Then the theory can be defined on any curved space-time as a topological quantum field theory, where the physical observables ${\cal O}$ are $Q_{A}$-closed operators (modulo the $Q_{A}$-exact relations). Hence we have a correspondence between the A-twisted NLSM and the geometric quantities of the target
\begin{equation}\label{}
  \begin{tabular}{|c|c|c|}
    \hline
    ${\rm NLSM}$ & $\leftrightarrow$ & ${\rm Target}$  \\
    $Q_{A}$ & $\leftrightarrow$ & $d$  \\
     $\text{span}\{\text{all }{\cal O}\}$ & $\leftrightarrow$ & ${\rm H}^{\ast}(X)$ \\
    \hline
  \end{tabular},
\end{equation}
where $d$ is the exterior derivative which can be used as the differential (coboundary) to define de Rham cohomology on the target. In this way, the data of quantum cohomology are encoded in the path-integral formulation as
\begin{equation}\label{ACF}
  \langle {\cal O}\rangle:=\int[D\Phi]{\cal O}e^{-S}.
\end{equation}
In this context, quantum cohomology is fairly well-understood through the two-dimensional nonlinear sigma model. We expect that quantum K theory can be investigated similarly if we perform a K-theoretic lifting of the 2d NLSM to the 3d NLSM defined over the spacetime $\Sigma\times S^{1}$, for any Riemann surface $\Sigma$ where NLSM lives. However, a general study of this theory is still missing. On the other hand, it was first observed by Witten that the NLSM can be regarded as the effective theory of a fundamental theory defined in the UV called the linear sigma model. The linear sigma model admits a straightforward 3d extension, namely the 3d Chern-Simons-matter theory. Many exact results of these supersymmetric gauge theories are known, so we focus mainly on gauge theory in this paper.\\

\noindent{\textbf{Gauged linear sigma model (GLSM)}}. By viewing the target space $X = \mathrm{Gr}(k;N)$ as the GIT-quotient $\mathbb{C}^{N}//U(k)$, one can readily accept that $X$ can be embedded into a $U(k)$ gauge theory with $N$ fields in the fundamental representation. We denote the vector multiplet by $V$, its superfield strength by $\Sigma$, and the $N$ matters by $\{\Phi_{i}\}_{i=1}^N$. The Lagrangian is
\begin{equation}\label{LSM}
  S=\int d^{2}xd^{4}\theta\sum^{N}_{i=1} {\rm Tr}{\bar \Phi}_{i}e^{V}\Phi_{i}+ \int d^{2}xd^{2}\widetilde{\theta}(-\frac{1}{2}t{\rm Tr}\Sigma)+c.c.,
\end{equation}
where $t=r-i\theta$ has real part $r$ being the FI-parameter associated with the $U(1)$ center of the $U(k)$ gauge group, and imaginary part $\theta$ being the theta angle of the gauge theory. The semi-classical ground state can be obtained by solving the vanishing of the potential energy U, which gives us the Grassmannian target. Furthermore, after integrating out the massive modes around the vacuum configuration, the theory will reduce to the NLSM on Grassmannian. Finally, the instanton moduli space of the gauge theory naturally provides a compactification (by ``quasi-maps'') to the moduli space of maps in the NLSM, and the mirror map (to ``stable maps'') can be understood as an RG-flow for \textit{point-like} instantons \cite{Witten:1993yc, Morrison:1994fr}. On the other hand, the path-integral of A-twisted genus-zero two-dimensional gauge theory can be computed exactly via the supersymmetric localization \cite{Benini:2015noa,Closset:2015rna} as
\begin{eqnarray}\label{AGCF1}
  \langle {\cal O }(\sigma|m)\rangle_{S^{2}}&=&\frac{1}{k!}\sum^{k}_{a=1}\sum^{\infty}_{l_{a}=0}\oint\prod^{k}_{a=1}\frac{d\sigma_{a}}{2\pi i}{\cal O }(\sigma|m)\left((-1)^{k-1}q\right)^{\sum^{k}_{a=1}l_{a}}\\\nonumber
  &&\prod_{1\leq a<b\leq k}\left(\sigma_{a}-\sigma_{b}\right)^{2}\prod^{N}_{i=1}\prod^{k}_{a=1}\left(\frac{1}{\sigma_{a}-m_{i}}\right)^{l_{a}+1}.
\end{eqnarray}
where ${\cal O }(\sigma|m)$ represents a general insertion, i.e., a polynomial
in $\sigma$'s and $m$'s. One can sum over all instantons to obtain a more concise formula
\begin{equation}\label{AGCF}
  \langle {\cal O }(\sigma|m)\rangle_{S^{2}}=\sum_{{\rm Vacua} }{\cal O }(\sigma|m)\left(\frac{\prod_{a<b}\left(\sigma_{a}-\sigma_{b}\right)^{2}}{\prod^{N}_{i=1}\prod^{k}_{a=1}\left(\sigma_{a}-m_{i}\right)}\left[\det_{a,b}\left(\frac{\partial^{2}\widetilde{W}^{2d}_{{\rm eff}}}{\partial \sigma_{a}\partial \sigma_{b}}\right)\right]^{-1}\right),
\end{equation}
where $\sigma_{a}$ is the lowest component field of the superfield strength $\Sigma_{a}$, and the two-dimensional twisted effective superpotential for this case is given by
\begin{equation}\label{2DE}
  \widetilde{W}^{2d}_{{\rm eff}}=-\left(t+i(k-1)\pi\right)\sum^{k}_{a=1}\Sigma_{a}-\sum^{N}_{i=1}\sum^{k}_{a=1}\left(\Sigma_{a}-m_{i}\right)\left(\log\left(\Sigma_{a}-m_{i}\right)-1\right).
\end{equation}
The vacuum equations are
\begin{equation}\label{VE}
  \exp\left(\frac{\partial \widetilde{W}^{2d}_{{\rm eff}}(\sigma_{a},m_{i})}{\partial\sigma_{a}}\right)=1,\quad\quad {\rm for}\quad a=1,\ldots,k,
\end{equation}
which gives
\begin{equation}\label{2VEGR}
  \prod_{i}\left(\sigma_{a}-m_{i}\right)=(-1)^{k-1}q,\quad\quad {\rm for} \quad a=1,\ldots,k, \quad\sigma_{a}\neq \sigma_{b}, \quad {\rm modulo}\ S_{k}.
\end{equation}
From the equations of motion of GLSM, one can find the geometric meaning of these $\sigma_{a}$'s: they are the Chern-roots of the tautological $U(k)$-bundle ${\cal S}$ over $X$. If we forget about the Weyl symmetry $S_{k}$ and the dynamical constraint, Eq.(\ref{2VEGR}) will give the equivariant quantum cohomology ring relations of $(\mathbb{CP}^{N-1})^{\times k}$. In this way, the nonabelian/abelian correspondence of quantum cohomology appears naturally in gauge theory. The pairing and structure constants can be computed from Eq.(\ref{AGCF}). Using the method developed initially in \cite{Gu:2020oeb}, one can also extract the quantum cohomology ring relations from the vacuum equations (\ref{VE}). To see this, we notice that there is a characteristic polynomial equation for Eq.(\ref{2VEGR}) where $\sigma_a\ (a=1,\cdots,k)$ give $k$ of the roots of:
\begin{equation}\label{CHP}
  \tau^{N} + \sum^{N}_{i=1}(-1)^i e_{i}(m)\tau^{N-i}=(-1)^{k-1}q.
\end{equation}
Here by convention, the elementary symmetric polynomials
\begin{equation}\label{}\nonumber
e_i(m) = \sum_{j_1< \cdots <j_i} m_{j_1}\cdots m_{j_i}.
\end{equation}
The above structure suggests that one should use the well-known Vieta formula to obtain relations among roots of the characteristic polynomial, which gives
\begin{eqnarray} \label{EQG}
   e_{1}(\sigma)+e_{1}\left(\widetilde{\sigma}\right)&=& e_{1}(m) \\\nonumber
  e_{2}(\sigma)+e_{1}(\sigma) e_{1}\left(\widetilde{\sigma}\right)+e_{2}\left(\widetilde{\sigma}\right)&=& e_{2}(m) \\\nonumber
  \vdots   \\\nonumber
  e_{k}(\sigma)e_{N-k}(\widetilde{\sigma}) &=& e_{N}(m)+(-1)^{k}q.
\end{eqnarray}
Here $\widetilde{\sigma} = \{\widetilde{\sigma}_a\}_{a=1}^{N-k}$ denotes the $(N-k)$ roots of Eq.(\ref{CHP}) other than $\sigma$, but we may also understand these as the collection of Chern roots of the quotient bundle
$\mathcal{Q}$.
%$\mathbb{C}^N/\mathcal{S}$.
The relations among the roots in Eq.(\ref{EQG}) give indeed the equivariant cohomology ring relations of Gr$(k;N)$. One can actually express the $e_{i}\left(\widetilde{\sigma}\right)$ by solving the first $N-k$ relations in Eq.(\ref{EQG}). More precisely,
\begin{equation}\label{SOEQG}
  e_{i}\left(\widetilde{\sigma}|m\right) =(-1)^{i}h_{i}\left(\sigma|m\right)=\sum^{i}_{j=1}(-1)^{i+j}h_{i-j}(\sigma)e_{j}(m).
\end{equation}
Therefore, each $e_{i}\left(\widetilde{\sigma}|m\right)$ is a specialization of the factorial homogenous polynomial of Chern-roots with all equivariant parameters located in $e_{j}(m)$, and is thus $S_{N}$-invariant. These factorial homogenous polynomials constitute a natural set of generators
%basis
for the equivariant quantum cohomology studied in \cite{Mihalcea:2008} and references therein. Plug Eq.(\ref{SOEQG}) into Eq.(\ref{EQG}), one can recover the presentation of the equivariant quantum cohomology of Grassmannian in Example \ref{Ex:1}. These properties of topological A-twisted gauge theory are enough for our purposes in this paper. Finally, we comment that all above can be obtained from the B-twisted nonabelian mirrors \cite{Gu:2018fpm}. \\

\noindent{\textbf{3d ${\cal N}=2$ Chern-Simons-matter theories}}. In three-dimensional gauge theory, one can also write down the Chern-Simons terms into the Lagrangian, in addition to the existing ones from the two-dimensional GLSMs. It is well known that the CS interaction affects the topological aspects of the theory. In this paper, we choose the spacetime to be $S^{2}\times S^{1}$ in the 3d ${\cal N}=2$ Chern-Simons-matter theories in order to focus on their quantum K theoretic aspects. 
The CS-interactions are
\begin{eqnarray}\label{CSA}
  S_{{\rm CS}}&=&\int_{S^{2}\times S^{1}}\left(\frac{\kappa_{SU(k)}}{4\pi i}{\rm Tr}\left(A\wedge dA-\frac{2i}{3}A^{3}\right)+\frac{{\kappa_{U(1)}- \kappa_{SU(k)}}}{4\pi i k}{\rm Tr}A\wedge d{\rm Tr}A\right.\\\nonumber
  &&\quad\quad\quad\quad\quad\quad +\left.\frac{\kappa_{R}}{4\pi i}{\rm Tr}A\wedge dA^{R}+\sum^{N}_{i=1}\frac{\kappa_{i}}{4\pi i}C^{i}\wedge d{\rm Tr}A+\ldots\right),
\end{eqnarray}
%\begin{eqnarray}\label{CSA}
%  S_{{\rm CS}}&=&\int_{S^{2}\times S^{1}}\left(\frac{\kappa_{SU(k)}}{4\pi i}{\rm Tr}\left(A\wedge dA-\frac{2i}{3}A^{3}\right)+\frac{k_{U(1)}-k_{SU(k)}}{4\pi i k}{\rm Tr}A\wedge d{\rm Tr}A\right.\\\nonumber
%  &&\quad\quad\quad\quad\quad\quad +\left.\frac{\kappa_{R}}{4\pi i}{\rm Tr}A\wedge dA^{R}+\sum^{N}_{i=1}\frac{\kappa_{i}}{4\pi i}C^{i}\wedge d{\rm Tr}A+\ldots\right),
%\end{eqnarray}
where $A$ is the $U(k)$ gauge field, $A^{R}$ is the background gauge field for the $U(1)$ R-symmetry, and $C^{i}$ is the background gauge field for the maximal flavor torus symmetry $T$. The omitted terms come from the ``super-partner'' of the CS interactions. The level parameters $\kappa_{U(k)}$, $\kappa_{R}$\footnote{This level is used for obtaining the Todd class of the target that appears in the GRR-formula.}, $\kappa_{i}$ are integers or half integers, and represent the bare gauge-CS level, the gauge-R-symmetry mixed CS level, and the flavor-gauge mixed CS level respectively. One may also consider other mixed CS interactions (see \cite{Jockers:2019lwe,Ueda:2019qhg}), but they affect the vacuum structure only by modifying the pairing \cite{Ueda:2019qhg}. The structure constants, however, only depend on $\kappa_{U(k)}$ and $\kappa_{i}$.

In CS-matter theory, the physical observables are Wilson loops associated with representations ${\cal R}$ of $U(k)$
\begin{equation}\label{3dPO}
  W_{{\cal R}}= {\rm Tr}_{{\cal R}}{\rm P}\exp(-\oint_{S^{1}}\sigma).
\end{equation}
They preserve half supersymmetry and are compatible with the topological A-twist on $S^2$. For example, considering a fundamental representation, we have
\begin{equation}\nonumber
  W_{\square}= {\rm Tr}_{\square}{\rm P}\exp(-\oint_{S^{1}}\sigma)=\sum^{k}_{a=1}x_{a}, \quad\quad x_{a}=e^{-2\pi R\sigma_{a}},
\end{equation}
where $R$ is the radius of $S^{1}$, and $\sigma_{a}=\frac{1}{2\pi R}\oint_{S^{1}}\sigma_{a}$ by abuse of notation. We denote by $t_{i}$ the equivariant parameters. Note, however, that their connection to the cohomological theory is not as direct as $t_i = e^{-2\pi R m_{i}}$, as we will see later.

In three-dimensional CS-matter theory, the K theoretic twisted effective superpotential is given by
\begin{eqnarray}\label{3dTS}\nonumber
  \widetilde{W}^{3d}_{{\rm eff}}&=&\frac{2\kappa_{SU(k)}+N}{4}\sum^{k}_{a=1}\left(\ln x_{a}\right)^{2}+\frac{\kappa_{U(1)}-\kappa_{SU(k)}}{2k}\left(\sum^{k}_{a=1}\ln x_{a}\right)^{2}+\left(\log(-1)^{k-1}q_{3d}\right)\sum^{k}_{a=1}\log x_{a}\\
  &&+\sum^{N}_{i=1}\sum^{k}_{a=1}\left(\kappa_{i}+\frac{1}{2}\right)\log t_{i}\log x_{a}+\sum^{N}_{i=1}\sum^{k}_{a=1}{\rm Li}_{2}(x_{a}t^{-1}_{i}).
\end{eqnarray}
The vacuum equations are
\begin{equation}\label{3DVE}
  1=\exp\left(\frac{\partial\widetilde{W}^{3d}_{{\rm eff}}}{\partial\log x_{a}}\right)=(-1)^{k-1}q_{3d}x^{\kappa_{SU(k)}+\frac{N}{2}}_{a}\prod^{k}_{b=1}x^{\frac{\kappa_{U(1)}-\kappa_{SU(k)}}{k}}_{b}\prod^{N}_{i=1}\frac{t^{\kappa_{i}+\frac{1}{2}}_{i}}{t_{i}-x_{a}}.
\end{equation}
There are
%infinite
infinitely many choices of gauge-invariant
%choices of
bare Chern-Simons levels. One first choice, as is investigated in \cite{Kapustin:2013hpk}, is
\begin{equation}\label{VER}
  \kappa_{U(1)}=\kappa_{SU(k)}=-\frac{N}{2},\quad\quad \kappa_{i}=\frac{1}{2}.
\end{equation}
In such case, the vacuum equations become
\begin{equation}\label{VEVER}
  \prod^{N}_{i=1}\left(1-x_{a}t^{-1}_{i}\right)=(-1)^{k-1}q_{3d}.
\end{equation}
Under the change of variables  $ x_a = 1 - e^{-L \sigma_a}, t_i = e^{-L m_i}$,
\begin{equation}\label{3m2}
  1-x_{a}t^{-1}_{i}\mapsto L(\sigma_{a}-m_{i}),\quad\quad q_{3d}\mapsto L^{N}q,
\end{equation}
where $L$ is the perimeter of $S^{1}$. They reduce to the vacuum equations of the equivariant quantum cohomology of $\mathrm{Gr}(k;N)$ as before. The mathematical theory can be found in \cite{Ruan2018}.

However, we focus mainly on the second choice of Chern-Simons levels here
\begin{equation}\label{GLQK}
    \kappa_{U(1)}=-\frac{N}{2},\quad\quad\kappa_{SU(k)}=k-\frac{N}{2},\quad\quad \kappa_{i}=\frac{1}{2},
\end{equation}
which gives Givental-Lee's quantum K theory. Indeed, under this choice, the vacuum equations become
\begin{equation}\label{GLVE}
    \prod^{N}_{i=1}(1-x_{a}t^{-1}_{i})=(-1)^{k-1}q_{3d}\frac{x^{k}_{a}}{\prod^{k}_{b=1}x_{b}}.
\end{equation}
Because of the extra factor on the right-hand side of the above equations, they are
different from the vacuum equations for $\left(\mathbb{CP}^{N-1}\right)^{\times k}$,
which explains in part why the nonabelian/abelian correspondence is not as
straightforward in Givental-Lee's quantum K theory. We will provide the
solution in section \ref{QKQHC}. For now, we comment only that the extra
factor in vacuum equations is natural in physics. For simplicity, we consider
the 3d CS-matter theory for the projective space $\mathbb{CP}^{N-1}$.
If we choose the Chern-Simons level $\kappa_{U(1)}=-\frac{N}{2}$,
the non-equivariant vacuum equation is
\begin{equation}\label{VEP1}
    (1-x)^{N}=q_{3d}.
\end{equation}
On the other hand, if we regard the projective space $\mathbb{CP}^{N-1}$ as $\mathbb{CP}^{N}[1]$, a degree-one hypersurface in $\mathbb{CP}^{N}$, and choose the level $\kappa_{U(1)}=-\frac{N+2}{2}$, the vacuum equation of gauge theory becomes
  \begin{equation}\label{VEP2}
    (1-x)^{N}=q_{3d}x^{-1}.
  \end{equation}
This is certainly not Givental-Lee's quantum K ring relation of $\mathbb{CP}^{N-1}$. However, if we choose a different level $\kappa_{U(1)}=-\frac{N}{2}$, we will get back the same vacuum equation (\ref{VEP1}). The physical reason is that one should choose a different level for the R-charge two matter. In nonabelian gauge theory, the roots of the gauge field can be regarded as R-charge two chiral matters \cite{Gu:2021yek,Gu:2023tcv}, so their levels should be different from matters with vanishing R-charges. It is reflected by the difference between $\kappa_{U(1)}$ and $\kappa_{SU(k)}$.

The correlation functions of these physical observables can be computed exactly by supersymmetric localization \cite{Benini:2015noa,Closset:2016arn,Ueda:2019qhg}
\begin{eqnarray}\label{3DCF}\nonumber
  \left \langle {\cal O}(x|t) \right\rangle^{\mathrm{QK}} &=& \sum^{\infty}_{l=0} \frac{\left((-1)^{k-1}q_{3d}\right)^{l}}{k!}\sum_{l_{1},\cdots,l_{k}\geq 0 \ \sum^{k}_{a=1}l_{a}=l}\sum^{N}_{i_{1},\cdots,i_{k}=1}\oint_{x_{a}=t_{i_{a}}}\prod^{k}_{a=1}\frac{dx_{a}}{2\pi i\left(x_{a}\right)^{k}}{\cal O}(x|t)\\
   &&\times  \prod_{1\leq a\neq b\leq k}\left(x_{b}-x_{a}\right)\prod^{N}_{i=1}\prod^{k}_{a=1}\left(\frac{t_{i}}{t_{i}-x_{a}}\right)^{l_{a}+1}\left(\prod^{k}_{a=1}\left(x_{a}\right)^{k\cdot l_{a}-l}\right).
\end{eqnarray}
{The vacuum equations (\ref{GLVE}) can also be derived from the above formula}. As we have done in the case of two-dimensional gauge theory, summing over the instanton moduli space, the above expression reduces to
\begin{eqnarray}\label{3DCF2}\nonumber
 \left \langle {\cal O}(x|t) \right\rangle&=&\sum_{{\rm Vacua}} {\cal O}(x|t)\left(\prod^{N}_{i=1}t^{\frac{k}{2}}_{i}\right)\left(\prod_{1\leq a< b\leq k}\left(x_{a}-x_{b}\right)^{2}\right)\left(\prod^{k}_{a=1}x^{\frac{N}{2}-(k-1)-\kappa_{R}}_{a}\right)\\
 &&\times\prod^{N}_{i=1}\prod^{k}_{a=1}\left(t_{i}-x_{a}\right)^{-1}\left[\det_{a,b}\frac{\partial^{2}\widetilde{W}^{3d}_{{\rm eff}}}{\partial\log x_{a}\partial\log x_{b}}\right]^{-1}.
\end{eqnarray}
With the known results reviewed above, we are now ready to present our new observation next, in section \ref{QKQHC}.

\section{A correspondence between the quantum K theory and the quantum cohomology}\label{QKQHC}
As mentioned in the previous sections, the nonabelian/abelian correspondence is formulated for the
quantum cohomology in the mathematical literature. However, no such techniques for quantum K theory are in the literature yet. In this paper, we shed light on this problem by showing that Givental-Lee's small quantum K theory can be regarded as a specialized equivariant small quantum cohomology which we call it the \textit{twisted quantum cohomology} (tQH).
\subsection{The non-equivariant situation}\label{NEC}
%To see this,
We first focus on the non-equivariant quantum K theory. The vacuum (or Bethe Ansatz) equations in physics are:
\begin{equation}\label{}\nonumber
  (1-x_{a})^{N}=(-1)^{k-1}q_{3d}\frac{x^{k}_{a}}{\prod^{k}_{b=1}x_{a}}.
\end{equation}
Following \cite{Gu:2022yvj}, we first symmetrize the above equations as
\begin{equation}\label{QKCP}
  \left(z_{a}\right)^{N}+\sum^{N-1}_{i=0}(-1)^{N-i}(z_{a})^{i}g_{N-i}(z_{a},q_{3d})=0,
\end{equation}
where $z_{a} = 1-x_{a}$. To state the formula for the polynomial $g_{N-i}(z_{a},q_{3d})$, following \cite{Gu:2020zpg,Gu:2022yvj}, we fix some notation. Set
\begin{equation}\label{}\nonumber
  c^{z}=\prod^{k}_{a=1}(1-z_{a})=\sum_{i\geq 0}(-1)^{i}e_{i}(z); \quad c^{z}_{\leq j}=\sum^{j}_{i=0}(-1)^{i}e_{i}(z); \quad c^{z}_{\geq j}=(-1)^{j}\left( c^{z}-c^{z}_{\leq j-1}\right).
\end{equation}
The polynomial coefficients $g_{N-i}(z_{a}, q_{3d})$ are given by
\begin{equation}\label{}\nonumber
  g_{i}(z_{a},q_{3d})=c^{z}_{\geq {i+1}}+(-1)^{N+k}{k-1 \choose N-i}q_{3d}\cdot\delta_{i,\geq N-k+1},
\end{equation}
where the notation $\delta_{i,\geq N-k+1}$ means that it is zero if $1\leq i\leq N-k$ and it is one if $N-k+1\leq i\leq N$. So the characteristic polynomial is
\begin{equation}\label{CPQK}
  \tau^{N}+\sum^{N-1}_{i=0}(-1)^{N-i}\tau^{i}g_{N-i}(z_{a},q_{3d})=0.
\end{equation}
Apply the Vieta formula to the above characteristic polynomial, we find that
\begin{eqnarray} \label{QKVF}
   e_{1}(z)+e_{1}\left(\widetilde{z}\right)&=& g_{1}(z_{a},q_{3d}) \\\nonumber
  e_{2}(z)+e_{1}(z) e_{1}\left(\widetilde{z}\right)+e_{2}\left(\widetilde{z}\right)&=& g_{2}(z_{a},q_{3d}) \\\nonumber
  \vdots   \\\nonumber
  e_{k}(z)e_{N-k}(\widetilde{z}) &=& g_{N}(z_{a},q_{3d}).
\end{eqnarray}
One can solve the first $N-k$ above equations, which gives us that
\begin{equation}\label{ESQ}
  e_{j}\left(\widetilde{z}\right)=(-1)^{j}{\cal O}_{j}=(-1)^{j}G_{j}.
\end{equation}

The characteristic polynomial (\ref{CPQK}) is quite different from the one for the non-equivariant quantum cohomology of Gr$(k;N)$:
\begin{equation}\label{}\nonumber
  \tau^{N}=(-1)^{k-1}q.
\end{equation}
However, we start from the condition that the characteristic polynomial of the
$G$-equivariant quantum cohomology is equal to that of the quantum K theory. To see the result,
%one can propose that the equivariant deformation of quantum cohomology could make the two characteristic %polynomials match. To see this,
we write the  {\em equivariant} quantum cohomology characteristic polynomial as
\begin{equation}\label{}\nonumber
  \prod^{N}_{i=1}\left(\tau-m_{i}\right)=(-1)^{k-1}q,
\end{equation}
which can be expanded as
\begin{equation}\label{DFE}
  \tau^{N}+\sum^{N-1}_{i=0}(-1)^{N-i}\tau^{i}e_{N-i}(m)+(-1)^{k}q=0.
\end{equation}
Since the equivariant parameters can be promoted to be fields in quantum field theory, if we perform the following
change of variables
% map
\begin{equation}\label{DM}
\begin{tabular}{|c|c|c|}
  \hline
  QK &  & $\mathrm{QH}^*_G$ \\
  $z_{a}$ & $\leftrightarrow$ & $L\sigma_{a}$ \\
  $q_{3d}$ & $\leftrightarrow$ & $L^Nq$ \\
   $g_{1}(z_{a},q_{3d})$ & $\leftrightarrow$ & $L e_{1}(m)$ \\
   & $\vdots$ &  \\
    $g_{N}(z_{a},q_{3d})$ & $\leftrightarrow$ & $L^{N} \left(e_{N}(m)+(-1)^{k}q\right)$  \\
  \hline
\end{tabular},
\end{equation}
then the two characteristic polynomials Eq.(\ref{CPQK}) and Eq.(\ref{DFE}) are identical.
We will make this more precise next. Consider the non-equivariant quantum K ring $\mathrm{QK}(\Gr(k;N))$.
A presentation for this ring was obtained in \cite{Gu:2022yvj}:
\begin{equation}\label{}\nonumber
  {\rm K}(pt)[\![q_{3d}]\!]\left[e_{1}(z),\cdots,e_{k}(z), e_{1}(\widetilde{z}),\cdots, e_{N-k}(\widetilde{z})\right]/\langle\sum_{i+j=\ell}e_{i}(z)e_{j}(\widetilde{z})-g_{\ell}\left(z,q_{3d}\right):1\leq\ell\leq N \rangle
\end{equation}
Consider now the $G=\GL(N)$-equivariant quantum cohomology ring $\QH^*_G(\Gr(k;N))$.
Observe that $H^*_G(pt)=\mathbb{Z}[m_1, \ldots, m_N]^{S_N}$. A variant of Witten's presentation for this ring is
\[
\QH^*_G(\Gr(k;N)) = \frac{\mathbb{Z}[q]\left[e_{1}(\sigma),\cdots,e_{k}(\sigma), e_{1}(\widetilde{\sigma}),\cdots, e_{N-k}(\widetilde{\sigma}), e_1(m), \ldots, e_N(m) \right]}{\left\langle\sum_{i+j=\ell}e_{i}(\sigma)e_{j}(\widetilde{\sigma})-e_{\ell}(m)-(-1)^{k}\delta_{\ell N} q : 1\leq \ell \leq N \right\rangle}
\]
The considerations above imply that there is a well-defined ring homomorphism
\[ \Phi: \QH^*_G(\Gr(k;N)) \to \QK(\Gr(k;N)) \]
defined by sending
\begin{equation}\label{E:defPhi} q \mapsto q_{3d}\/; \quad e_i(\sigma) \mapsto e_i(z) \/; \quad e_j(\tilde{\sigma}) \mapsto e_j(\tilde{z}) \/; \quad
e_\ell(m)+(-1)^{k}\delta_{\ell N}q \mapsto g_\ell(z;q_{3d}) \/, \end{equation}
for $1 \le i \le k, 1 \le j \le N-k, 1 \le \ell \le N$.
For simplicity of notation, we made the homogenization variable $L =1$; this will be added later on.

{An alternate algebraic construction is as follows. Consider the ideal
$J \subset \QH^*_G(\Gr(k;N))$
defined by
\[ J = \langle e_\ell(m)+(-1)^{k}\delta_{\ell N}q - g_\ell(\sigma;q); 1 \le \ell \le N \rangle \/. \]
(In other words, this ideal encodes the relations \eqref{DM}.)
Clearly, $J$ is included in the kernel of $\Phi$. Define the {twisted quantum cohomology (tQH)  ring}:
\begin{equation}\label{E:ATQH-def} \widetilde{\QH}^\ast_G(\Gr(k;N)) = \QH^\ast_G(\Gr(k;N)) / J \/. \end{equation}
Then $\Phi$ induces a ring homomorphism
\[ \widetilde{\Phi}: \widetilde{\QH}^\ast_G(\Gr(k;N)) \to \mathrm{QK}(\Gr(k;N) \]
defined by the assignments in \eqref{E:defPhi}. We expect that $\widetilde{\Phi}$ is an
algebra isomorphism.}

In quantum K theory, the classes of the structure sheaves of Schubert varieties give a natural basis for the quantum K ring.
%from the geometric point of view. However,
We will not review these details in this note and recommend the readers find the geometric definition in section 5.1 of \cite{Gu:2022yvj}. Furthermore, a quantum \textit{Giambelli formula} for the quantum K theory was proved in \cite{Buch:2008}, {see also \cite{Gorbounov:2014}.} It states that
%, which says that
every K-theoretic quantum Schubert class is a polynomial of the special classes ${\cal O}_{i}$, $1\leq i\leq N-k$. So we can restrict to those classes. Finally, we recall
%adopt another well-known fact
that the Grothendieck polynomials $G_{j}(z)$ give representatives of the structure sheaves
%of Schubert varieties
${\cal O}_{j}\in {\rm K}( {\rm Gr}(k;N))$. Since the Grothendieck polynomials are linear combinations
%is a linear combination
of Wilson loops in gauge theory, thus we can perform the computation by using tools in quantum field theory.

On the other hand, as discussed in \cite{Mihalcea:2008}, the factorial Schur functions, $S_{T}(\sigma|m)$, are the basis of the equivariant quantum cohomology of Grassmannain. A quantum Giambelli formula in this situation has been proved in \cite{Mihalcea:2008},
%which taught us that
giving the general Schur function $S_{T}(\sigma|m)$
%can be constructed from
as a determinant in the factorial homogenous polynomials $h_{j}\left(\sigma|m\right)$. %{\color{blue} Now, we are ready to show that the structure sheaves of Schubert varieties in quantum K theory are the factorial Schur functions in the specialized equivariant quantum cohomology under the map (\ref{DM}).}
%{\color{red} LM: Is this ever shown ? WG: Not explicitly shown before this. LM: Then I am not sure how to justify this statement. I suggest to remove the portion above in blue.WG: I have delated it.}
A connection between the factorial Schur functions and Grothendieck polynomials is obtained next. We first notice that the Grothendieck polynomials can be expressed in terms of the elementary functions and the homogenous functions as in \cite{Lenart:2000}:
\begin{equation}\label{GRP}
  G_{j}(z)=\sum_{a,b\geq 0, a+b\leq k}(-1)^{b}h_{j+a}(z)e_{b}(z).
\end{equation}
Under the dictionary \eqref{DM}, it is not difficult to show that
\begin{equation*}\label{}
 L^{j} h_{j}\left(\sigma|m\right)\mapsto G_{j}(z).
\end{equation*}
\begin{example} For ${\rm Gr}(k;N)$, we denote by ${\cal O}_{\lambda}$ the K-theory class for the structure sheaf of the Schubert variety corresponding to a Young tableaux $\lambda$. For ${\rm Gr}(2; 4)$, we have six Schubert classes: ${\cal O}_{\emptyset}, {\cal O}_{1}, {\cal O}_{1, 1}, {\cal O}_{2}$, ${\cal O}_{2,1}, {\cal O}_{2,2}$. They can be expressed in terms of Wilson loops as
\begin{eqnarray*}
   {\cal O}_{1} &=& 1-W_{1,1}, \\
  {\cal O}_{1,1} &=& 1-W_{1}+W_{1,1}, \\
  {\cal O}_{2} &=& 1-3W_{1,1}+W_{2,1}, \\
  {\cal O}_{2,1} &=& 1-W_{1}+W_{2,1}-W_{2,2}, \\
   {\cal O}_{2,2} &=& 1-2W_{1}+W_{2}+3W_{1,1}-2W_{2,1}+W_{2,2}.
\end{eqnarray*}
Following the dictionary, the map for ${\rm Gr}(2; 4)$ is
\begin{equation}\label{}
  L e_{1}(m)=e_{2}(z),\quad  L^{2} e_{2}(m)=0,\quad L^{3} e_{3}(m)=q_{3d},\quad e_{4}(m)=0.
\end{equation}
A direct computation says:
\begin{eqnarray} \nonumber
  L h_{1}(\sigma|m) &=& L h_{1}(\sigma)-Le_{1}(m)=h_{1}(z)-e_{2}(z)=G_{1}(z), \\\nonumber
  L^{2} h_{2}(\sigma|m) &=& L^{2} h_{2}(\sigma)-L^{2}h_{1}(\sigma)e_{1}(m)+L^{2}e_{2}(m)\\\nonumber
  &=&h_{2}(z)-e_{1}(z)e_{2}(z)=G_{2}(z).
\end{eqnarray}

\end{example}

A natural question is to find the analogue of the quantum K pairing under the above dictionary. This will be done next.
%So the question is: what does it map to in quantum cohomology?

%%%%%%%%%%%%%%%%%%%%%%%%%%%%%%%%%%%%%%%%%%%%%%%%%%%%%%%%%%
%%%%%%%%%%%%%%%%%%%%%%%%%%%%%%%%%%%%%%%%%%%%%%%%%%%%%%%%%

%\subsection{Associated Twisted Quantum Cohomology}\label{ATQH}
\subsection{Quantum $K$ theory via residues}\label{ATQH}
 In this section we find a formula for the pairing in the QK ring in
 terms of a residue calculation. This gives a conjectural statement based on physics 
 for the quantum K pairing. In this section we will work non-equivariantly.
 
 We start by recalling the non-equivariant correlation functions (\ref{3DCF}) for the
 quantum K theory of Grassmannian:
\begin{eqnarray}\label{neqkc}
  \left \langle {\cal O}(x_{a}) \right\rangle^{\mathrm{QK}} &=& \sum^{\infty}_{l=0} \frac{\left((-1)^{k-1}q_{3d}\right)^{l}}{k!}\sum_{l_{1},\cdots,l_{k}\geq 0 \/, \sum^{k}_{a=1}l_{a}=l}\oint_{x_{a}=1}\prod^{k}_{a=1}\frac{dx_{a}}{2\pi i\left(x_{a}\right)^{k}}{\cal O}(x_{a})\\\nonumber
   &&\times  \prod_{1\leq a\neq b\leq k}\left(x_{b}-x_{a}\right)\prod^{k}_{a=1}\left(\frac{1}{1-x_{a}}\right)^{N(l_{a}+1)}\left(\prod^{k}_{a=1}\left(x_{a}\right)^{k\cdot l_{a}-l}\right).
\end{eqnarray}
%{\color{red} LM: Change order of residue and products ?}
Here $ {\cal O}(x_{a}) $ is a symmetric polynomial in the variables $x_1, \ldots, x_k$, interpreted as exponentials of the Chern roots
of the tautological subbundle $\mathcal{S}$ of $\Gr(k;N)$ (of rank $k$). The residue is the 
Jefffrey-Kirwan residue e.g. from \cite{brion.vergne:JF}. 
A key property of the
correlation function is that it only depends on the image of the polynomial $ {\cal O}(x_{a}) $ in the quantum K ring $\mathrm{QK}(\Gr(k;N))$. 
Equivalently, the correlation function applied to any polynomial relation in the quantum K ring is equal to $0$. 
For example, one may deduce the correlator function applied to the vacuum/Bethe Ansatz equations \eqref{GLVE} 
is equal to $0$, and the relations in the quantum K ring are consequences of these equations.

We would like to calculate the quantum K correlation function by rewriting its defining residue in a form where we may utilize the (non-equivariant) quantum cohomology correlation function
from \eqref{AGCF1}. We first change the K-theoretic variables into cohomological ones:
\begin{equation}\label{E:cov-QK-QH} q_{3d} = L^n q_{2d} \/; \quad x_a = 1 - L \sigma_a \/; \quad \cO(x_a) = \cO(\sigma_a) \end{equation}
Next we observe that, when compared to \eqref{AGCF1}, there 
are two extra factors in the correlation functions of the quantum K theory of Gr$(k;N)$ (\ref{3DCF}).
The rational function
\begin{equation*}
 \left(\prod^{k}_{a=1}\left(x_{a}\right)^{k\cdot l_{a}-l}\right)=\left(\prod^{k}_{a=1}\left(1-L\sigma_{a}\right)^{k\cdot l_{a}-l}\right)
\end{equation*}
encodes the effect of the equivariant deformation parameters $m$ via the dictionary \eqref{DM}. 
(For the projective space, i.e. when $k=1$, this 
factor does not arise.) Another key difference from cohomology to K theory is given by the rational function:
\begin{equation}\label{E:keyfactor}
  \frac{1}{\prod^{k}_{a=1} \left(x_{a}\right)^{k}}=\frac{1}{\prod^{k}_{a=1} \left(1-L\sigma_{a}\right)^{k}}
\end{equation}
Geometrically, this arises as $1/ \det (\mathcal{S})^{\star k}$, where $\star$ denotes the quantum $K$ multiplication.
It was proved in \cite{Gu:2022yvj} that if $\mathcal{Q}$ denotes the tautological quotient bundle on $\Gr(k;N)$ (of rank $N-k$), then
\[ \det \mathcal{S} \star \det \mathcal{Q} = 1-q_{3d} \/. \]
Furthermore, in the ordinary K-theory ring, the (Poincar{\'e}) dual basis of the Schubert basis $\{ \cO^\lambda \}$ consists of the
elements $\{ \det \mathcal{S} \cdot \cO^{\lambda^\vee} \}$, where $\lambda^\vee$ is the complement of $\lambda$ in the 
$k \times (N-k)$ rectangle; see \cite{qkchevalley}. It follows that in $K(\Gr(k;N))$, 
\[ \det \mathcal{Q} = \sum_\lambda \cO_\lambda \/.\]
(See the next section for precise definitions of the Schubert classes $\cO_\lambda$.) 
Finally, in the quantum K ring the Schubert classes $\cO^\lambda$ are represented by Grothendieck polynomials 
$G_\lambda(x_1, \ldots, x_k)$; see \cite{Gorbounov:2014}.
Combining all of this implies that in the calculation of the quantum K correlation function \eqref{neqkc}
we may replace the factor $\frac{1}{\prod^{k}_{a=1} \left(x_{a}\right)^{k}}$ above by
$\frac{ (\det \mathcal{Q})^{\star k}}{(1-q_{3d})^k}$. In the quantum K ring, this expression 
is represented by 
\[ \frac{ (\sum_\lambda G_\lambda(\sigma))^k}{(1-q_{3d})^k} \]
After utilizing \eqref{E:cov-QK-QH} to transform everything into cohomological variables, 
and absorbing the extra Grothendieck polynomials into the factor $\cO(\sigma)= \cO(\sigma_1, \ldots, \sigma_k)$,  
we will need to calculate the `twisted quantum cohomology' (tQH)
correlation function:
\begin{equation}\label{E:tQK-cor}
\begin{split}  \langle \cO(\sigma) \rangle^{\mathrm{tQH}} & = \sum^{\infty}_{l=0} \frac{\left((-1)^{k-1}L^{N}q_{2d}\right)^{l}}{k!}\sum_{l_{1},\cdots,l_{k}\geq 0 \/, \sum^{k}_{a=1}l_{a}=l}\oint_{\sigma_{a}=0}\prod^{k}_{a=1}\frac{-d(L\sigma_{a})}{2\pi i}{\cal O}(\sigma_a)\\ 
   & \times  \prod_{1\leq a\neq b\leq k}\left(L\sigma_{a}-L\sigma_{b}\right)\prod^{k}_{a=1}\left(\frac{1}{L\sigma_{a}}\right)^{N(l_{a}+1)}\left(\prod^{k}_{a=1}\left(1-L\sigma_{a}\right)^{k\cdot l_{a}-l}\right) \/. \end{split} 
 \end{equation}
Informally, this means that
\begin{equation}\label{E:maintQH} \left\langle \frac{\cO(\sigma)}{\det (\mathcal{S})^{\star k}} \right\rangle^{\mathrm{tQH}} = \langle \cO(x) \rangle^{\mathrm{QK}} \/,\end{equation}
i.e., the QK correlation function may be calculated from the twisted one, after one replaces $1/(\det \mathcal{S})^{\star k}$
by an appropriate polynomial representative in $\mathrm{QK}(\Gr(k;N))$. The terminology `twisted quantum cohomology' is supported by the fact that this is equal to the quantum cohomology correlation
function \eqref{AGCF1} multiplied by the last factor $\prod (1 - L \sigma_a)^{k l_a -l}$.   
The classical case (i.e., when $l=0$) does not involve this factor, and a direct proof of \eqref{E:maintQH} will be discussed in the Appendix \ref{CGP}.

If one utilizes that $\det \mathcal{S} \star \det (\mathcal{Q}) =1 - q_{3d}$ and the previous discussion,
one obtains that 
\begin{equation}\label{E:maineq}  \langle \cO(x) \rangle^{\mathrm{QK}} = 
\frac{1}{(1-q_{3d})^k} \left\langle \cO(\sigma) \cdot \left(\sum_\lambda G_\lambda(\sigma)\right)^k \right\rangle^{\mathrm{tQH}} \/.\end{equation}
By linearity, this gives an expression for the quantum K pairing in terms of the twisted correlator function for quantum 
K structure constants and a single Grothendieck polynomials. More precisely, write
\[ \cO_\lambda \star \cO_\mu \star (\det \mathcal{Q})^{\star k} = \sum_{\nu,l} C_{\lambda,\mu,k}^{\nu,d} q_{3d}^l \cO_\nu \]
in $\QK(\Gr(k;N))$. Then by the definition of the quantum K pairing via the correlators, we have:
\[ \langle{\cal O}_{\lambda}, {\cal O}_{\mu} \rangle  
=  \langle G_{\lambda}(\sigma) \cdot G_{\mu}(\sigma) \rangle^{{\rm QK}} 
= \frac{1}{(1-q_{3d})^k}\sum_{\nu,l} C_{\lambda,\mu,k}^{\nu,l}q_{3d}^l \langle G_\nu(\sigma) \rangle^{\mathrm{tQH}} \/. \]
 Therefore a natural question is how to calculate the one-point correlators for Grothendieck polynomials
 $G_\nu(\sigma)$ indexed by a partition $\nu=(\nu_1, \ldots, \nu_k)$ included in the $k \times (N-k)$ rectangle. 
 It is known that $G_\nu(\sigma)$ is a symmetric polynomial, with the maximal power with respect to 
 any of the individual variables $\sigma_i$ at most
$\nu_1 \le N-k$; see e.g., \cite[Thm. 2]{Lenart:2000}. We claim that the only non-vanishing contribution in \eqref{E:tQK-cor} comes from $l=0$. To see this,
w.l.o.g. we may assume that $k\leq N-k$, and that $l_{1}\geq l_{a}$ ($a \ge 2$), with $l\geq 1$ .
The residue with respect to the indeterminate $\sigma_{1}$ is the coefficient in front of
$\sigma_{1}^{-1}$. We calculate the power of $\sigma_1$ as 
\[ (N-k+2(k-1)+kl_{1}-l)-N(l_{1}+1) \] 
where $N-k$, $2(k-1)$, $kl_{1}-l$ 
are the maximal powers of $\sigma_{1}$ from the terms in
$G_{\nu}$, $\prod_{1\neq b\leq k}\left(\sigma_{1}-\sigma_{b}\right)^{2}$, and $\left(1-L\sigma_{1}\right)^{k\cdot l_{1}-l}$, respectively. Observe that 
\[ N(l_{1}+1)-1 - (N-k+2(k-1)+kl_{1}-l) = (N-k)l_1 +l - k +1 \ge l+1 >1 \]
%Since
%\begin{equation*}
%  N(l_{1}+1)-1>N-k+2(k-1)+kl_{1}-l = N+k + k l_1 -l-2 \/,
%\end{equation*}
thus the coefficient of $\sigma_1^{-1}$ is equal to zero. Therefore the sought residue is equal to  
\begin{eqnarray}\label{utc2}
  \left \langle G_{\nu}(\sigma) \right\rangle^{{\rm tQH}} = \frac{1}{k!}\oint_{\sigma_{a}=0}\prod^{k}_{a=1}\frac{-d(L\sigma_{a})}{2\pi i}G_{\nu}(\sigma)
   \times  \prod_{1\leq a\neq b\leq k}\left(L(\sigma_{a}-\sigma_{b})\right)\prod^{k}_{a=1}\left(\frac{1}{L\sigma_{a}}\right)^N \/.
\end{eqnarray}
%{\color{red}which will not be modified under the equivariant deformation, and this means that it also equals to 
%\begin{eqnarray}\label{utc3}
%% \nonumber to remove numbering (before each equation)
%%\frac{1}{k!}
%\sum_{1 \le i_1 < i_2 < \ldots < i_k \le N} \oint_{\sigma_{a}=m_{i_a}}\prod^{k}_{a=1}\frac{-d(L\sigma_{a})}{2\pi i}G_{\nu}(\sigma)
%   \times  \prod_{1\leq a\neq b\leq k}\left(L(\sigma_{a}-\sigma_{b})\right) \prod_{j=1}^N\prod^{k}_{a=1}\left(\frac{1}{L\sigma_{a}-Lm_{j}}\right).
%\end{eqnarray}}
By the nonabelian/abelian correspondence theorem proved in \cite{Martin:2000} or \cite[section 9]{KOUY}, this residue is equal to the ordinary 
cohomological integral 
\begin{equation}\label{E:topform} top(G_\nu) := \int_{\Gr(k;N)} G_\nu(\sigma) \/, \end{equation} 
i.e., the coefficient of the Schur function
$s_{(N-k)^k}(\sigma)$ in $G_{\nu}(\sigma)$. (Equivalently, this is the coefficient of the $\sigma$-homogeneous part of 
degree $k(N-k) = \dim \Gr(k,N)$.) We refer to this as the `top-form integral', or simply, the
`top-form', and denote it by $top(G_\nu)$. In other words, we have shown that 
\begin{equation}\label{E:topform-tQK} \left\langle G_\nu(\sigma) \right\rangle^{\mathrm{tQH}} = top(G_\nu) \/. \end{equation}
{In Conjecture \ref{conj:pairing} in the next section we will give a different (conjectural) interpretation 
of the top form, in terms of an Euler characteristic in the quantum K theory ring. This will give an expression for the 
residue defining the tQH correlator for any degree $d$.}

%A mathematical consequence of this calculation will be stated in the Conjecture \ref{conj:pairing} below. %in section \ref{AMFPQK}%below.

 Unfortunately, a direct connection between the (twisted) quantum cohomology and quantum
K theory correlation functions for $l\neq 0$ is unclear. We take a moment to explain the difficulty.
Consider the quantum cohomology nonabelian/abelian correspondence for ${\rm Gr}(k;N)$ as in 
\cite[\S 12]{KOUY}, see also \cite{Martin:2000,BCK05}. Let ${\cal O}\left(\sigma\right)$ be a 
symmetric polynomial in the 
Chern roots $\sigma_1, \ldots, \sigma_k$, interpreted 
as an element in $H^*(X)$. 
By equations (12.1.4) and (12.2.1) in \cite{KOUY}, 
an integral over the degree $l=\sum^{k}_{a=1} l_{i}$ moduli 
space $\overline{{\cal M}}_{l}$ of quasimaps to $\Gr(k,N)$ may be calculated as:
\begin{equation}\label{E:QH-ANA}
  \int_{\overline{{\cal M}}_{l}} {\cal O}\left(\sigma\right)=\frac{(-1)^{(k-1)l}}{k!}\sum_{d=\sum^{k}_{i=1} l_{i}}\int_{\prod^{k}_{a=1}\left(\mathbb{P}^{N(1+l_{a})-1}\right)}\prod_{b\neq c}\left(\sigma_{b}-\sigma_{c}\right) {\cal O}\left(\sigma\right).
\end{equation}
(We refer to \cite{KOUY} for relevant details.) 
There is a (multi)residue formula for calculating the integral on the right hand side.
On the projective space
\begin{equation*}
  \int_{\mathbb{P}^{N-1}}{\cal O}(\sigma)=\oint_{\sigma=0}\frac{d\sigma}{2\pi i\sigma^{N}}{\cal O}(\sigma).
\end{equation*}
Combining the two equations, we arrive at
\begin{equation*}
   \int_{\overline{{\cal M}}_{l}} {\cal O}\left(\sigma\right)=\frac{{(-1)^{(k-1)l}}}{k!}\sum_{d=\sum^{k}_{i=1} l_{i}} \left(\oint_{\sigma_a=0} \prod^{k}_{a=1} \frac{d\sigma_{a}}{2\pi i \left(\sigma_{a}\right)^{(N+1)l_{a}}}\right)\prod_{a\neq b}\left(\sigma_{b}-\sigma_{c}\right) {\cal O}\left(\sigma\right).
\end{equation*}
However, if $l>0$ this residue differs from the twisted residue \eqref{E:tQK-cor} by the twisting factor $\prod (1-L \sigma_a)^{k l_a-l}$. The main question is to find a (hopefully easy to formulate) relation between the two residues. Defining a correlation function for the tQH ring in \eqref{E:ATQH-def}
along with a study of the nonabelian/abelian correspondence on this ring may shed light on this question. Earlier evidence supporting a quantum K theoretic nonabelian/abelian
correspondence  comes from a calculation of the K-theoretic $J$ function by Taipale \cite{Taipale:2013}.

\begin{example} We calculate $\langle 1 \rangle^{\mathrm{QK}}$ for $\Gr(2,4)$, utilizing \eqref{E:maineq} and 
\eqref{E:topform-tQK}. Using that $\det \mathcal{Q}= \sum \cO_\lambda$ and the calculations from \cite{Buch:2008}, or \cite[eqs.(2.76)-(2.90)]{Gu:2020zpg},
we have:
\begin{eqnarray}\label{QGR24}
 \left(\det {\cal Q}\right)^{\star 2}  &=& (1+4q_{3d}+q^{2}_{3d})+(2+4q_{3d}){\cal O}_{1}+(3+3q_{3d}){\cal O}_{1,1} \\
   &&  +(3+3q_{3d}){\cal O}_{2}+(5+q_{3d}){\cal O}_{2,1}+6{\cal O}_{2,2},
\end{eqnarray}
% where we have used the quantum structure constants of the Schubert basis, cf.~e.g.\cite[Eq.(2.76) to Eq.(2.90)]{Gu:2020zpg}. 
The non-zero contributions to the top form
 from \eqref{E:topform-tQK} in $\left(\det {\cal Q}\right)^{\star 2}$ only arise from
 \begin{equation}\label{}\nonumber
   (5+q_{3d})G_{2,1}(\sigma)+6G_{2,2}(\sigma)=(6-5-q_{3d})\left(\sigma_{1}\sigma_{2}\right)^{2}
   +(5+q_{3d})\sigma_{1}\sigma_{2}\left(\sigma_{1}+\sigma_{2}\right).
 \end{equation}
Notice that $\langle \left(\sigma_{1}\sigma_{2}\right)^{2}\rangle^{{{\rm tQH}}}=1$ and 
$\langle \sigma_{1}\sigma_{2}\left(\sigma_{1}+\sigma_{2}\right)\rangle^{{{\rm tQH}}}=0$, 
{because $(\sigma_{1}\sigma_{2})^2$ is the Schur function $s_{(2,2)}(\sigma_1,\sigma_2)$,
and that $\sigma_1 \sigma_2 (\sigma_1+\sigma_2) = \sigma_{(2,1)}(\sigma_1, \sigma_2)$.}  
As an illustration of the Conjecture \ref{conj:pairing} below, we obtain:

\begin{equation*}
  \langle 1 \rangle^{\mathrm{QK}} = \frac{\oint 1 \star \det \mathcal{Q}^{\star 2}}{(1-q_{3d})^2}=\frac{1-q_{3d}}{(1-q_{3d})^2} =\frac{1}{1-q_{3d}}.
\end{equation*}
A similar calculation applies to other pairings to give:

\begin{equation}\label{}\nonumber
\left(\left\langle G_\lambda(x) \cdot G_\mu(x) \right\rangle^{{\rm QK}}\right)_{\lambda,\mu}=\left(
  \begin{array}{cccccc}
    \frac{1}{1-q_{3d}} & \frac{1}{1-q_{3d}} & \frac{1}{1-q_{3d}} & \frac{1}{1-q_{3d}}& \frac{1}{1-q_{3d}}& \frac{1}{1-q_{3d}}\\
    \frac{1}{1-q_{3d}} & \frac{1}{1-q_{3d}} & \frac{1}{1-q_{3d}} & \frac{1}{1-q_{3d}}& \frac{1}{1-q_{3d}}& \frac{q_{3d}}{1-q_{3d}}\\
     \frac{1}{1-q_{3d}} & \frac{1}{1-q_{3d}} & \frac{1}{1-q_{3d}} & \frac{q_{3d}}{1-q_{3d}}& \frac{q_{3d}}{1-q_{3d}}&\frac{q_{3d}}{1-q_{3d}}\\
   \frac{1}{1-q_{3d}} & \frac{1}{1-q_{3d}} & \frac{q_{3d}}{1-q_{3d}} & \frac{1}{1-q_{3d}}& \frac{q_{3d}}{1-q_{3d}}& \frac{q_{3d}}{1-q_{3d}}\\
   \frac{1}{1-q_{3d}} & \frac{1}{1-q_{3d}} & \frac{q_{3d}}{1-q_{3d}} & \frac{q_{3d}}{1-q_{3d}}& \frac{q_{3d}}{1-q_{3d}}& \frac{q_{3d}}{1-q_{3d}}\\
   \frac{1}{1-q_{3d}} & \frac{q_{3d}}{1-q_{3d}} & \frac{q_{3d}}{1-q_{3d}} & \frac{q_{3d}}{1-q_{3d}}& \frac{q_{3d}}{1-q_{3d}}& \frac{\left(q_{3d}\right)^{2}}{1-q_{3d}}\\
  \end{array}
\right).
\end{equation}
This is consistent with the mathematical calculations of the QK pairing from \cite{Buch2020,Buch:2008}. 
\end{example}
More examples may be found in the next section. 

\subsection{A relation to the mathematical quantum K pairing}\label{AMFPQK}
 In \cite{Buch2020}, a formula for pairing of the quantum K ring of any (generalized) flag variety has been found. We recall this formula when $X= \mathrm{Gr}(k;N)$ is a
Grassmannian. Fix the (opposite) standard flag $F_1 \subset F_2 \subset \ldots \subset F_N = \mathbb{C}^N$, where 
$F_i= \langle e_N, \ldots , e_{N-i+1} \rangle$. 
For each partition $\lambda =(\lambda_1\ge \ldots \ge \lambda_k)$ such that
$\lambda_k \ge 0$ and $\lambda_1 \le N-k$, define the Schubert variety
\[ X_\lambda = \{ V \in \mathrm{Gr}(k;N): \dim V \cap F_{N-k+i-\lambda_i} \ge i \} \/. \]
This is a subvariety of complex codimension $|\lambda| = \lambda_1 + \ldots + \lambda_k$. We denote
its fundamental class by $\sigma_\lambda :=[X_\lambda] \in H^{2 |\lambda|}(\mathrm{Gr}(k;N))$ and
by $\cO_\lambda =[\cO_{X_\lambda}] \in K(\mathrm{Gr}(k;N))$ the (Grothendieck) class
determined by the structure sheaf of the Schubert variety.

Let $\lambda, \lambda'$ be two partitions included in the
$k \times (N-k)$ rectangle, and let $d(\lambda,\lambda')$ be the minimum
degree in the quantum cohomology multiplication of
$\sigma_\lambda \star \sigma_{\lambda'}$.
It was proved in \cite{Buch2020} that the QK pairing is given by:
\[ \langle \cO_\lambda, \cO_{\lambda'}\rangle = \frac{q^{d(\lambda, \lambda')}}{1-q_{3d}} \/. \]
Next, we relate this formula with the pairing (\ref{E:maineq}) obtained from physics considerations.

For $a \in \mathrm{K}(X)$ define the `top form integral' $\oint_{X}^{H} a$
as follows. Expand $a$ into Schubert classes: $a = \sum_{\lambda \subset (N-k)^k} a_\lambda \mathcal{O}_\lambda$,
then consider the associated polynomial \[ \sum {a_{\lambda} G_\lambda (\sigma)} \/,\]
where $G_{\lambda }(\sigma)= G_{\lambda} (\sigma_1, \ldots, \sigma_k)$
is the Grothendieck polynomial. Then
\[ \oint_{X}^{H} a { :=} \sum a_{ \lambda} top(G_{ \lambda}) \/,\]
where recall from \eqref{E:topform-tQK} that the top-form $top(G_{\lambda})$ is defined to be the coefficient of the Schur function $s_{(N-k)^k}(z)$ in $G_{\lambda}(\sigma)$.

%Consider also the usual K-theoretic pairing 

%We will refer to
%$top(G_\lambda(z_1, \ldots, z_k))$ as the {\em top form} of $G_\lambda$.}
Then the statement (\ref{E:maineq}) for quantum K pairings can be restated as follows.

\begin{conj}\label{conj:pairing} Consider any elements $a_1, \ldots, a_p \in K(X)$. Then
the correlator for the monomial $a_1\cdot \ldots \cdot a_p$ is equal to:   
\[ \langle a_1\cdot \ldots \cdot a_p \rangle^{tQH} = \frac{\chi_{\Gr(k;N)} \bigl( a_1 \star \ldots \star a_p \star (\det \mathcal{S})^{\star k}\bigr)}{1-q_{3d}}\/, \]
where $\chi_{\Gr(k;N)}$ is extended by $q$-linearity.

Furthermore, for any $a \in K(X)$, the top form of $a$ is given by
\[ \oint_{\Gr(k;N)}^H a = \frac{\chi_{\Gr(k;N)} (a \star (\det \mathcal{S})^{\star k})}{1-q_{3d}} \/. \]
\end{conj}
A proof of this conjecture in the classical case, i.e., when $q=0$, is given in the Appendix. It uses a residue calculation on the `abelianization' of $\mathrm{Gr}(k;N)$, namely $(\mathbb{P}^{N-1})^k$.

 A corollary of this conjecture are two formulae for calculating the quantum K pairing. On one side, for $a,b \in K(X)$, 
\[ \langle a, b \rangle = \left< \frac{ a \star  b }{ (\det \cS)^{\star k}} \right>^{\rm {tQH}} =
\frac{ \chi (a \star b) }{1-q_{3d}} \/. \]
The right hand side of this formula is implicit in \cite{Buch2020}. 
A new formula may be obtained
by calculating the tQH correlator directly, and using that $ \det \cS \star \det \cQ = 1-q_{3d}$:
\[  \langle a, b \rangle = \frac{ \oint_{\Gr(k;N)} a \star  b \star (\det \cQ)^{\star k}}{(1-q_{3d})^k} \/. \]

To illustrate the conjecture in the simplest case, consider the particular case when $\lambda=\lambda'=\emptyset$. In this case,
according to \cite{Buch2020}, $\langle 1,1\rangle = \frac{1}{1-q_{3d}}$. Then the first part of the conjecture states that
\[ \frac{1}{1-q_{3d}}= \left< \frac{1}{(\det \mathcal{S})^{\star k}} \right>^{{\rm tQH}} =\frac{\chi_{\Gr(k;N)} (\cO_{\Gr(k;N)})}{1-q_{3d}} \]
Since $\det \mathcal{S} \star \det \mathcal{Q} =1 -q_{3d}$, the last part of the conjecture may be restated as a remarkable equality involving a top form:
\[ \left< (\det \mathcal{Q})^{k} \right>^{{\rm tQH}}= \oint^{H}_X \det(\mathcal{Q})^{\star k} = (1-q_{3d})^{k-1} \/. \]

\begin{example}\label{ex:PN} Consider $k=1$, i.e., $X$ is the complex projective space $\mathbb{P}^{N-1}$. The Schubert
classes are indexed by partitions $(i)$ for $0 \le i \le N-1$. For projective spaces, the quantum K ring multiplication coincides with the quantum cohomology multiplication \cite[Thm. 5.4]{Buch:2008}:
\[ \cO_i \star \cO_j = \begin{cases} \cO_{i+j} & i + j < N \\
q \cO_{i+j-N} & i+j \ge N \/.\end{cases}\]
We have that $\det \mathcal{Q} = 1 + \cO_1 + \ldots + \cO_{N-1}$ and the Grothendieck polynomials
are $G_i(z_1) =z_1^i$, for $0 \le i \le N-1$. It follows that $\oint^{QH}_X \cO_i = \delta_{i,N-1}$. One may check that the coefficient
of $\cO_{N-1}$ in $\cO_i \star \det \mathcal{Q}$ is always equal to $1$, which implies
that the coefficient of $\cO_{N-1}$ in $\cO_i \star \cO_j \star \det \mathcal{Q}$ is equal to $1$ if
$i+j < N$ and $q$ otherwise. This proves the conjecture in this case.
\end{example}

\begin{example} We further illustrate the conjecture in some examples in $\mathrm{Gr}(3,6)$. Recall that $\det(\mathcal{Q}) = \sum_{\lambda \subset (3,3,3)} \cO_\lambda$.
A computer calculation shows that
\[ \begin{split} \det & (\mathcal{Q})^{\star 2}= (q^3+9q^2+9q+1)
+ (6q^2+12q+2)\cO_1\\
& +(4q^2+13q+3)\cO_2+(4q^2+13q+3)\cO_{1,1} +(4q^2+12q+4)\cO_{1,1,1}\\
& +(4q^2+12q+4)\cO_3+(q^2+14q+5)\cO_{2,1}+(q^2+12q+
7)\cO_{2,1,1}\\
&+(q^2+12q+7)\cO_{3,1}+(14q+6)\cO_{2,2}+(q^2+9q+10)\cO_{3,1,1}+(11q+9)
\cO_{2,2,1}
\\ & +(11q+9)\cO_{3,2}+(10q+10)\cO_{2,2,2}+(6q+14)\cO_{3,2,1}+(10q+10)\cO_{3,3} \\
&+(4q+16)\cO_{3,2,2}+(4q+16)\cO_{3,3,1}+(q+19)\cO_{3,3,2}+20\cO_{3,3,3}
\end{split}
\]
and that
\[ \begin{split} \det&(\mathcal{Q})^{\star 3} =
(118q^3+227q^2+54q+1)+(50q^3+245q^2+102q+3)\cO_{1}
\\ &+(32q^3+224q^2+138 q+6)\cO_{2}+(32q^3+224q^2+138q+6)\cO_{1,1}\\
&+(28q^3+216q^2+146q+10)\cO_{1,1,1}+(
28q^3+216q^2+146q+10)\cO_{3}\\
&+(12q^3+172q^2+202q+14)\cO_{2,1}+(7q^3+153q^2+
215q+25)\cO_{2,1,1}\\&
+(7q^3+153q^2+215q+25)\cO_{3,1}+(12q^3+138q^2+230q+20)\cO_{
2,2}\\
&+(q^3+126q^2+227q+46)\cO_{3,1,1}
+(7q^3+108q^2+245q+40)\cO_{2,2,1}\\
&+(7q^3+108q^2+245q+40)\cO_{3,2}+(7q^3+98q^2+245q+50)\cO_{2,2,2}
\\&
+(q^3+64q^2+251q+84)
\cO_{3,2,1}
+(7q^3+98q^2+245q+50)\cO_{3,3}\\
&+(q^3+50q^2+239q+110)\cO_{3,2,2}+(q^3+
50q^2+239q+110)\cO_{3,3,1}\\
&+(q^3+35q^2+209q+155)\cO_{3,3,2}+(q^3+35q^2+189q+
175)\cO_{3,3,3}
\end{split}
\]
The Grothendieck polynomials $G_\lambda=G_\lambda(z_1, z_2, z_3)$ which have non-zero top form have
$\lambda \supset (3,2,1)$ and are given by
\begin{equation}\label{E:non-eq36} top(G_{3,2,1}) = -1; \quad top(G_{3,2,2}) =top(G_{3,3,1}) = top(G_{3,3,3}) =1 \/; \quad
top(G_{3,3,2}) = -2 \/. \end{equation}
One calculates that
\[ \begin{split} \oint_{\mathrm{Gr}(3,6)}^{H} \det(\mathcal{Q})^{\star 3}  = &
 -(q^3+64q^2+251q+84)+(q^3+50q^2+239q+110)
\\ &+(q^3+ 50q^2+239q+110) -2 (q^3+35q^2+209q+155)
\\ & + (q^3+35q^2+189q+175)
\\ &=  q^2 - 2q +1\\ & =  (1-q)^2 \end{split}\]
This confirms the conjecture for $\lambda=\lambda'=\emptyset$.\end{example}

Further examples illustrating the conjecture, in the more general equivariant case, may be found in the next section. We have checked this conjecture for all Grassmannians $\mathrm{Gr}(k;N)$ for $N \le 7$. The observant reader may have noticed the fact that all the coefficients in the expansions of
$\det \mathcal{Q}^{\star k}$ are positive; this is surprising, given that the Schubert quantum K structure constants have signs alternating with the sum of codimensions \cite{BCMP:qkpos}.

% 
% {\color{blue} including giving an explanation of the positivity of the coefficients arising the quantum K multiplication by $\det(\mathcal{Q})$.}

\subsection{The equivariant situation}\label{NC}
The vacuum equations of the equivariant quantum K ring are
\begin{equation}\label{EVEQ}
 \left( \prod^{N}_{i=1}\left(z_{a}-r_{i}\right)\right)\left(\prod_{b\neq a}\left(1-z_{b}\right)\right)+(-1)^{k}q_{3d}(1-z_{a})^{k}\prod^{N}_{i=1}(1-r_{i})=0,
\end{equation}
where $r_{i}=1-t_{i}$. The symmetrization of Eq.(\ref{EVEQ}) has been investigated in \cite{Gu:2022yvj}:
\begin{equation}\label{SCP}
 (z_{a})^{N}+\sum^{N-1}_{i=0}(z_{a})^{i}g_{N-i}(z_{a}, r_{j}, q_{3d})=0.
\end{equation}
The polynomial
\begin{equation} \label{E:charpoly}
 \tau^{N}+\sum^{N-1}_{i=0}\tau^{i}g_{N-i}(z_{a}, r_{j}, q_{3d})=0
\end{equation}
is called the charactetistic polynomial; see \cite{Gu:2022yvj}.
To represent the coefficients $g_{j}(z_{a}, r_{i}, q_{3d})$, we adopt the notation $c^{r}$, which has a similar structure as $c^{z}$ used in section \ref{NEC}. We further set
\begin{equation}\label{}\nonumber
  c^{\prime}_{\geq\ell}\left(z,r\right)=e_{\ell}(r)+e_{\ell-1}(r)c^{z}_{\geq 2}+e_{\ell-2}(r)c^{z}_{\geq 3}+\ldots+e_{\ell-k+1}(r)c^{z}_{\geq k},
\end{equation}
when $r_{i}=0$, we have $c^{\prime}_{\geq\ell}\left(z,0\right)=c^{z}_{\geq\ell+1}$ by using the usual convention that $e_{i}(z)=e_{i}(r)=0$ for $i<0$ and $e_{0}(z)=e_{0}(r)=1$. Define the matrices
\begin{eqnarray*}
  E &=& \left(
     \begin{array}{cccc}
       -1 & 0 & \ldots & 0 \\
       -e_{1} & -1 & \ldots & 0 \\
       \vdots & \vdots & \ddots & 0 \\
       -e_{k-1} & -e_{k-2} & \ldots & -1 \\
     \end{array}
   \right); \\
   C^{r}_{\geq N-k+2}&=&\left(
                        \begin{array}{c}
                          c^{r}_{\geq N-k+2} \\
                          \vdots \\
                          c^{r}_{\geq N} \\
                          0 \\
                        \end{array}
                      \right); \quad\quad
   C^{z,r}_{\geq N-k+1}\ =\ \left(
                             \begin{array}{c}
                               c^{\prime}_{\geq N-k+1} \\
                               c^{\prime}_{\geq N-k+2} \\
                               \vdots \\
                               c^{\prime}_{\geq N} \\
                             \end{array}
                           \right).
\end{eqnarray*}
Following \cite{Gu:2022yvj}, the polynomials $g_{j}(z_{a}, r_{i}, q_{3d})$ are equal to:
%Then we are ready to
%express the polynomial coefficients $g_{j}(z_{a}, r_{i}, q_{3d})$ as
\begin{equation}\label{}\nonumber
  \begin{cases}
  c^{\prime}_{\geq\ell}\left(z,r\right) & \text{if  $1\leq \ell\leq N-k$ }\\
  c^{\prime}_{\geq\ell}\left(z,r\right)+\left(E\cdot C^{r}_{\geq N-k+2}\right)_{\ell}+(-1)^{N+k}q_{3d}{k-1 \choose N-\ell}c^{r} & \text{if  $N-k+1\leq \ell\leq N$ }.
  \end{cases}
\end{equation}
%%The above calculation is a tedious, but rather standard algebraic manipulation.

The updated map between the equivariant quantum K theory and the equivariant quantum cohomology is
  \begin{equation}\label{DM2}
\begin{tabular}{|c|c|c|}
  \hline
  ${\rm QK^*_G}$ &  & ${\rm QH^*_{\widetilde{G}}}$ \\
  $z_{a}$ & $\leftrightarrow$ & $L\sigma_{a}$ \\
   $g_{1}(z_{a}, r_{i}, q_{3d})$ & $\leftrightarrow$ & $L e_{1}(\hat{m})$ \\
   & $\vdots$ &  \\
    $g_{N}(z_{a}, r_{i}, q_{3d})$ & $\leftrightarrow$ & $L^{N} \left(e_{N}(\hat{m})+(-1)^{k}q\right)$  \\
  \hline
\end{tabular},
\end{equation}
where $\hat{m}$ is the twisted mass (or equivariant parameter) in quantum cohomology. We use a different notation for the twisted mass to match the equivariant quantum K theory.
In other words, this is the same as \eqref{DM} except that the polynomials
$g_i(z_{a}, r_{i}, q_{3d})$ are now updated to the equivariant setting.
We expect that there is a ring isomorphism
\[ {\rm QH}^{*}_G({\rm Gr}(k;N))/J^G \simeq \mathrm{QK}_G(\Gr(k;N)) \]
where the ideal
$J^G=
   \langle e_{\ell}(\hat{m})+(-1)^k\delta_{\ell N}q-g_{\ell}(z_{a}, r_{i}, q_{3d}); 1\leq \ell\leq N\rangle$
and which sends
\[ q \mapsto q_{3d}\/; \quad e_i(\sigma) \mapsto e_i(z) \/; \quad e_j(\tilde{\sigma}) \mapsto e_j(\tilde{z}) \/; \quad
e_\ell(\hat{m})+(-1)^{k}\delta_{\ell N}q \mapsto g_\ell(z;r_i;q_{3d}) \/, \]
for $1 \le i \le k, 1 \le j \le N-k, 1 \le \ell \le N$.
As usual $\sigma$ and $\widetilde{\sigma}$ denote the equivariant Chern roots of the tautological bundles
$\mathcal{S}$ and $\mathcal{Q}$ respectively, and $z=(z_1, \ldots, z_k), \tilde{z}=(\tilde{z}_1, \ldots, \tilde{z}_{n-k})$ denote the roots of the characteristic polynomial \eqref{E:charpoly}.
We also made the homogenization variable $L=1$.

%
%So we find that the equivariant quantum K theory is isomorphic to the ring
%%equivariant quantum cohomology
%${\rm \widetilde{QH}}^{*}_G({\rm Gr}(k;N))$ defined by
%\begin{equation*}
%  {\rm \widetilde{QH}}^{*}_G({\rm Gr}(k;N))={\rm QH}^{*}_G({\rm Gr}(k;N))/J,
%\end{equation*}
% where the ideal $J$ is given by
% \begin{equation*}
%   \langle e_{\ell}(\widetilde{m})+(-1)^k\delta_{\ell N}q-g_{\ell}(z_{a}, r_{i}, q_{3d}); 1\leq \ell\leq N\rangle.
% \end{equation*}
%{\color{red} LM: Very unclear. What equivariant quantum cohomology ring are we talking about ? Presumably again $QH^*_G(Gr(k,N))$ with $G= GL_N$?
%If yes, what is the precise statement giving the relation to this ring ?}
%
It was proved in \cite{Gu:2022yvj} that in the presentation above of the equivariant quantum K theory, the elementary symmetric functions satisfy
\begin{equation}\label{}
  e_{j}(\widetilde{z})=(-1)^{j}G^{\prime}_{j}(z,t)=(-1)^{j}\sum_{a+b=j}(-1)^{a}e_{a}(t)G_{b}(z)\/.
\end{equation}
Here $G'_j(z,t)$ is a symmetric version of the factorial Grothendieck polynomial from \cite{McNamara:2005}.
%is far from being the factorial Grothendieck polynomial \cite{McNamara:2005}.
%The geometric explanation for $G^{\prime}_{j}(z,r)$ is still missing. Nevertheless,
%One can show that under the dictionary:
%\begin{equation*}%\label{}
  %h_{j}(\sigma|\widetilde{m}) \mapsto G^{\prime}_{j}(z,r)
  %\mapsto L^{j} h_{j}(\sigma|\widetilde{m}),
%\end{equation*}
the structure constants in terms of the basis $G^{\prime}_{\lambda}(z,t)$ are identical to the ones in equivariant cohomology. Furthermore, the statement about correlation functions extends to the equivariant setting
in the same way, meaning that:
%deformations is the same as for the non-equivariant case. This means
\begin{equation}\label{}
  \left\langle\prod_{j}G^{\prime}_{\lambda^j}(z_{a},t) \right\rangle^{{\rm {QK}}} = \left\langle \prod_{\lambda^j}G^{\prime}_{\lambda^j}(\sigma_{a},t) \left(\det\left(1-L\sigma_{a}\right)\right)^{-k} \right\rangle^{{\rm {\mathrm{tQH}}}}.
\end{equation}
{Of course, one can insert more general classes on both sides by extending the discussion in section \ref{ATQH} to the equivariant situation. Let us first recall that the correlators (\ref{3DCF})
\begin{eqnarray}\label{eqkc} \nonumber
  \left \langle {\cal O}(x|t) \right\rangle^{\mathrm{QK}} &=& \sum^{\infty}_{l=0} \frac{\left((-1)^{k-1}q_{3d}\right)^{l}}{k!}\sum_{l_{1},\cdots,l_{k}\geq 0 \ \sum^{k}_{a=1}l_{a}=l}\sum^{N}_{i_{1},\cdots,i_{k}=1}\oint_{x_{a}=t_{i_{a}}}\prod^{k}_{a=1}\frac{dx_{a}}{2\pi i\left(x_{a}\right)^{k}}{\cal O}(x|t)\\
   &&\times  \prod_{1\leq a\neq b\leq k}\left(x_{b}-x_{a}\right)\prod^{N}_{i=1}\prod^{k}_{a=1}\left(\frac{t_{i}}{t_{i}-x_{a}}\right)^{l_{a}+1}\left(\prod^{k}_{a=1}\left(x_{a}\right)^{k\cdot l_{a}-l}\right),
\end{eqnarray} 
 which motivates us to define the so-called equivariant `twisted quantum cohomology' 
correlation function:
\begin{equation}\label{E:etQK-cor}
\begin{split}  \langle \cO(\sigma|m) \rangle^{\mathrm{tQH}} & = \sum^{\infty}_{l=0} \frac{\left((-1)^{k-1}L^{N}q_{2d}\right)^{l}}{k!}\sum_{l_{1},\cdots,l_{k}\geq 0 \/, \sum^{k}_{a=1}l_{a}=l}\oint_{\sigma_{a}=0}\prod^{k}_{a=1}\frac{-d(L\sigma_{a})}{2\pi i}{\cal O}(\sigma|m)\\ 
   & \times  \prod_{1\leq a\neq b\leq k}\left(L\sigma_{a}-L\sigma_{b}\right)\prod^{N}_{i=1}\prod^{k}_{a=1}\left(\frac{1-Lm_{i}}{L\sigma_{a}-Lm_{i}}\right)^{l_{a}+1}\left(\prod^{k}_{a=1}\left(1-L\sigma_{a}\right)^{k\cdot l_{a}-l}\right) \/. \end{split} 
 \end{equation}
 This also says
\begin{equation} \label{E:mainetQH} \left\langle \frac{\cO(\sigma|m)}{\det (\mathcal{S})^{\star k}} \right\rangle^{\mathrm{tQH}} = \langle \cO(x|t) \rangle^{\mathrm{QK}} \/,\end{equation}
which is an equivariant version of Eq.(\ref{E:maintQH}). Now, we shall utilize $\det \mathcal{S} \star \det \mathcal{Q} = (1-q_{3d})\prod_{i=1}^N t_i$, then Eq.(\ref{E:mainetQH}) will be replaced by
\begin{equation}
\label{E:mainetQH2}  \langle \cO(x|t) \rangle^{\mathrm{QK}}= \frac{1}{\left((1-q_{3d})\cdot \prod_{i=1}^N t_i\right)^k}\left\langle \cO(\sigma|m)\cdot \det (\mathcal{Q})^{\star k}\right\rangle^{\mathrm{tQH}} \/.\end{equation}
 }

We now turn to the equivariant version of the QK and twisted QH correlators. The main definitions and facts will closely match the non-equivariant situation. 
We continue to use the notation 
\[x_i = e^{-L \sigma_i}, \quad t_j = e^{-L m_j} \]
where $\sigma_i$ ($1 \le i \le k$) are the (equivariant) 
Chern roots of $\mathcal{S}$ and $m_j$
$(1 \le j \le N)$ is the mass (or equivariant parameter) associated to the vector space 
spanned by the basis vector $e_j$. However, as is customary when dealing  with equivariant K theory classes, we do not distinguish between cohomological and K theoretic Chern roots. In particular, we abuse notation and denote 
\[ \sigma_i = 1 - x_i \/; \quad m_j = 1 - t_j \/. \]
We will use the definition of the double Grothendieck polynomials from McNamara \cite{McNamara:2005}, see also \cite[Prop. 2.3]{Gorbounov:2014}, and we recall next the basic facts we need about these polynomials. The double Grothendieck polynomials
are non-homogeneous 
polynomials $G_\lambda(\sigma| \tilde{m})$
in the variables $\sigma= (\sigma_1, \ldots, \sigma_k)$ above and 
$\tilde{m}=(\tilde{m}_1, \ldots, \tilde{m}_N)$, such that if one specializes the mass parameters 
$\tilde{m}_i=0$ then 
one obtains the ordinary Grothendieck polynomials 
$G_\lambda(\sigma)$. To harmonize with the conventions from \cite{Gorbounov:2014} 
we need to make 
\[ \tilde{m}_j = 1 - t_j^{-1} \]
As usual we define the degrees of the variables to be
\[ \deg \sigma_i = 1 \/; \quad \deg \tilde{m}_j = 1 \/. \]
With this definition,
\[ G_\lambda(\sigma | \tilde{m} ) = s_\lambda(\sigma | \tilde{m}) + \textrm{ higher order terms } \]
where the {\em factorial Schur polynomial} $s_\lambda(\sigma| \tilde{m})$ is 
by definition the homogeneous
part of $G_\lambda(\sigma |m )$ of degree $|\lambda| = \lambda_1 + \ldots + \lambda_k$. 

Fix $\lambda$ a partition included in the $k \times (N-k)$ rectangle. The {\em top form} of $G_\lambda(\sigma|\tilde{m})$, denoted by 
\begin{equation}\label{E:eqtopform}\oint_{\Gr(k;N)} G_\lambda(\sigma| \tilde{m}) \quad \in \mathbb{Z}[\tilde{m}_1, \ldots, \tilde{m}_N] \subset \mathrm{K}_{T}(pt) \/, \end{equation}
is defined to be the coefficient of $s_{(N-k)^k}(\sigma|\tilde{m})$ in the 
expansion of $G_\lambda(\sigma|\tilde{m})$.  
It follows from \cite[Thm. 2]{Lenart:2000} that this is the same as the coefficient of 
$s_{(N-k)^k}(\sigma)$
in $G_\lambda(\sigma|\tilde{m})$. (More precisely, this follows because the coefficient of $s_\nu(\sigma)$ in $G_\lambda(\sigma|\tilde{m})$
is nonzero only if $\nu$ is included in the $k \times \lambda_1$ rectangle. 
In the non-equivariant case, i.e., for $t_j=1$, a combinatorial formula for this coefficient is 
given in \cite{Lenart:2000}.) For reasons which will become more clear later we give the top form in terms of the variables $t_j^{-1}$ instead of $\tilde{m}_j=1-t_j^{-1}$.
\begin{example} Let $k=2, N=4$. Then the factorial Grothendieck polynomial 
\[ G_{(2,1)}(\sigma| \tilde{m}) = G_{(2,1)}(\sigma_1,\sigma_2|\tilde{m}_1, \tilde{m}_2, \tilde{m}_3, \tilde{m}_4) \] 
is equal to:
\begin{align*}
& -\tilde{m}_1^2 \tilde{m}_2 \tilde{m}_3 \sigma_1^2 \sigma_2^2 + 2 \tilde{m}_1^2 \tilde{m}_2 \tilde{m}_3 \sigma_1^2 \sigma_2 + 2 \tilde{m}_1^2 \tilde{m}_2 \tilde{m}_3 \sigma_1 \sigma_2^2 + \tilde{m}_1^2 \tilde{m}_2 \sigma_1^2 \sigma_2^2 \\
& + \tilde{m}_1^2 \tilde{m}_3 \sigma_1^2 \sigma_2^2 + 2 \tilde{m}_1 \tilde{m}_2 \tilde{m}_3 \sigma_1^2 \sigma_2^2 - \tilde{m}_1^2 \tilde{m}_2 \tilde{m}_3 \sigma_1^2 - 4 \tilde{m}_1^2 \tilde{m}_2 \tilde{m}_3 \sigma_1 \sigma_2 \\
& - \tilde{m}_1^2 \tilde{m}_2 \tilde{m}_3 \sigma_2^2 - 2 \tilde{m}_1^2 \tilde{m}_2 \sigma_1^2 \sigma_2 - 2 \tilde{m}_1^2 \tilde{m}_2 \sigma_1 \sigma_2^2 - 2 \tilde{m}_1^2 \tilde{m}_3 \sigma_1^2 \sigma_2 - 2 \tilde{m}_1^2 \tilde{m}_3 \sigma_1 \sigma_2^2 \\
& - \tilde{m}_1^2 \sigma_1^2 \sigma_2^2 - 3 \tilde{m}_1 \tilde{m}_2 \tilde{m}_3 \sigma_1^2 \sigma_2 - 3 \tilde{m}_1 \tilde{m}_2 \tilde{m}_3 \sigma_1 \sigma_2^2 - 2 \tilde{m}_1 \tilde{m}_2 \sigma_1^2 \sigma_2^2 - 2 \tilde{m}_1 \tilde{m}_3 \sigma_1^2 \sigma_2^2 \\
& - \tilde{m}_2 \tilde{m}_3 \sigma_1^2 \sigma_2^2 + 2 \tilde{m}_1^2 \tilde{m}_2 \tilde{m}_3 \sigma_1 + 2 \tilde{m}_1^2 \tilde{m}_2 \tilde{m}_3 \sigma_2 + \tilde{m}_1^2 \tilde{m}_2 \sigma_1^2 + 4 \tilde{m}_1^2 \tilde{m}_2 \sigma_1 \sigma_2 \\
& + \tilde{m}_1^2 \tilde{m}_2 \sigma_2^2 + \tilde{m}_1^2 \tilde{m}_3 \sigma_1^2 + 4 \tilde{m}_1^2 \tilde{m}_3 \sigma_1 \sigma_2 + \tilde{m}_1^2 \tilde{m}_3 \sigma_2^2 + 2 \tilde{m}_1^2 \sigma_1^2 \sigma_2 + 2 \tilde{m}_1^2 \sigma_1 \sigma_2^2 \\
& + \tilde{m}_1 \tilde{m}_2 \tilde{m}_3 \sigma_1^2 + 4 \tilde{m}_1 \tilde{m}_2 \tilde{m}_3 \sigma_1 \sigma_2 + \tilde{m}_1 \tilde{m}_2 \tilde{m}_3 \sigma_2^2 + 3 \tilde{m}_1 \tilde{m}_2 \sigma_1^2 \sigma_2 + 3 \tilde{m}_1 \tilde{m}_2 \sigma_1 \sigma_2^2 \\
& + 3 \tilde{m}_1 \tilde{m}_3 \sigma_1^2 \sigma_2 + 3 \tilde{m}_1 \tilde{m}_3 \sigma_1 \sigma_2^2 + 2 \tilde{m}_1 \sigma_1^2 \sigma_2^2 + \tilde{m}_2 \tilde{m}_3 \sigma_1^2 \sigma_2 + \tilde{m}_2 \tilde{m}_3 \sigma_1 \sigma_2^2 \\
& + \tilde{m}_2 \sigma_1^2 \sigma_2^2 + \tilde{m}_3 \sigma_1^2 \sigma_2^2 - \tilde{m}_1^2 \tilde{m}_2 \tilde{m}_3 - 2 \tilde{m}_1^2 \tilde{m}_2 \sigma_1 - 2 \tilde{m}_1^2 \tilde{m}_2 \sigma_2 - 2 \tilde{m}_1^2 \tilde{m}_3 \sigma_1 - 2 \tilde{m}_1^2 \tilde{m}_3 \sigma_2 \\
& - \tilde{m}_1^2 \sigma_1^2 - 3 \tilde{m}_1^2 \sigma_1 \sigma_2 - \tilde{m}_1^2 \sigma_2^2 - \tilde{m}_1 \tilde{m}_2 \tilde{m}_3 \sigma_1 - \tilde{m}_1 \tilde{m}_2 \tilde{m}_3 \sigma_2 - \tilde{m}_1 \tilde{m}_2 \sigma_1^2 - 4 \tilde{m}_1 \tilde{m}_2 \sigma_1 \sigma_2 \\
& - \tilde{m}_1 \tilde{m}_2 \sigma_2^2 - \tilde{m}_1 \tilde{m}_3 \sigma_1^2 - 4 \tilde{m}_1 \tilde{m}_3 \sigma_1 \sigma_2 - \tilde{m}_1 \tilde{m}_3 \sigma_2^2 - 3 \tilde{m}_1 \sigma_1^2 \sigma_2 - 3 \tilde{m}_1 \sigma_1 \sigma_2^2 \\
& - \tilde{m}_2 \tilde{m}_3 \sigma_1 \sigma_2 - \tilde{m}_2 \sigma_1^2 \sigma_2 - \tilde{m}_2 \sigma_1 \sigma_2^2 - \tilde{m}_3 \sigma_1^2 \sigma_2 - \tilde{m}_3 \sigma_1 \sigma_2^2 - \sigma_1^2 \sigma_2^2 + \tilde{m}_1^2 \tilde{m}_2 + \tilde{m}_1^2 \tilde{m}_3 \\
& + \tilde{m}_1^2 \sigma_1 + \tilde{m}_1^2 \sigma_2 + \tilde{m}_1 \tilde{m}_2 \sigma_1 + \tilde{m}_1 \tilde{m}_2 \sigma_2 + \tilde{m}_1 \tilde{m}_3 \sigma_1 + \tilde{m}_1 \tilde{m}_3 \sigma_2 + \tilde{m}_1 \sigma_1^2 + 2 \tilde{m}_1 \sigma_1 \sigma_2 + \tilde{m}_1 \sigma_2^2 \\
& + \tilde{m}_2 \sigma_1 \sigma_2 + \tilde{m}_3 \sigma_1 \sigma_2 + \sigma_1^2 \sigma_2 + \sigma_1 \sigma_2^2
\end{align*}
The factorial Schur function $s_{(2,2)}(\sigma|\tilde{m}) = \sigma_1^2 \sigma_2^2$ + other terms. 
After making changes of variables 
%$x_i = 1 - \sigma_i$ ($i=1, 2$) and 
$t_j^{-1} = 1-\tilde{m}_j$ ($j=1, \ldots , 4$), the coefficient of $s_{(2,2)}(\sigma)$ in $G_{(2,1)}(\sigma|t)$ is 
%the same as the coefficient of $x_1^2 x_2^2$
%in $G_{(2,1)}(\sigma|m)$, and it is 
equal to $-(t_1^{2} t_2t_3)^{-1}$, that is:
\[ \oint_{\Gr(2;4)} G_{(2,1)}(\sigma|t) = -(t_1^2 t_2t_3)^{-1} \/. \]
\end{example}
As usual, we extend the notion of top form to any class $a \in K_T(\Gr(k;N))$ by considering the expansion
$a = \sum_\lambda a_\lambda \cO_\lambda$, and defining
\[ \oint_{\Gr(k;N)} a = \sum_\lambda a_\lambda \oint_{\Gr(k;N)} G_\lambda \/. \]
 
%{\color{red} LM: We do not need this.}
%Again, the powers of
%$\det \cQ$ in the equivariant quantum K theory of $\Gr(k;N)$ will play a key role. 
%The expansion of this class in terms of Schubert classes is:
%\[ \det \cQ = \sum_\lambda  t_{j_{1}}\cdot \ldots \cdot t_{j_{N-k}} \cO_\lambda \]
%where $1 \le j_i \le N$ the position of the $i$-th horizontal step in the partition 
%$\lambda$, as read
%in the SW-NE order, in the $(N-k, \ldots, N-k)$ rectangle. 
%For example, if $k=3, N=7$ and $\lambda= (3,1)$ then
%\[ (1 \le j_1< j_2 < j_3< j_4 \le 7-3) = (2<4<5<7) \/. \]
%This calculation follows from the the calculation of duals of Schubert classes
%in the equivariant K theory ring $K_T(\Gr(k;N))$; see \cite[Prop. 4.2]{qkchevalley}. 

\begin{conj}\label{conj:pairing2} Let $a_1, \ldots , a_p \in K_T(\Gr(k;N))$. Then the tQH correlator is equal to 
\[ \langle a_1 \cdot \ldots \cdot a_p \rangle^{tQH} = \frac{\chi_{\Gr(k;N)}( a_1 \star \ldots \star a_p \star (\det \cS)^{ \star k} )}{1-q_{3d}}\/. \]
Furthermore, for any $a \in K_T(\Gr(k;N))$, the top form is related to the right hand side by:
\[ \oint_{\Gr(k;N)}a = \frac{\chi_{\Gr(k;N)}( a \star (\det \cS)^{ \star k})}{(t_1 \ldots t_N)^k(1-q_{3d})} \/. \]
\end{conj}
%Note that the quantity $(t_{w_\lambda(1)} \cdot \ldots \cdot t_{w_\lambda(k)})^k$ is also equal to the equivariant localization of $(\det \mathcal{S})^k$ (note this is the classical power) at the fixed point $\mathbf{e}_\lambda$. By the localization formula in the equivariant K theory ADDREF, this implies that for $a \in K_{T}(\Gr(k;N))$, 
%\begin{equation}\label{E:eqtopform} \oint_{\Gr(k;N)} a = \chi_{\Gr(k;N)} (a \otimes (\det \mathcal{S})^k ) \/. \end{equation}
%

\begin{example} We illustrate the equivariant conjecture for 
$\mathrm{Gr}(2,4)$. To start, we observe that $\oint_{\Gr(2,4)} \cO_\lambda = 0$ 
unless $\lambda \supset (2,1)$. In these cases we have:
\[ \oint_{\mathrm{Gr}(2,4)} \cO_{(2,1)} =-(t_1^{2} t_2  t_3)^{-1} \/; \quad
\oint_{\mathrm{Gr}(2,4)} \cO_{(2,2)} = (t_1^{2} t_2^{2})^{-1} \/. \]
Then, according to the conjecture:
\[ \langle \cO_{(2,1)} \rangle^{tQH} = \frac{\chi( \cO_{(2,1)} \star (\det \cS)^{\star 2})}{1-q_{3d}} \/. \]
To compare the statement about the top form, note that
$\det \cS= t_1 t_2(1- \cO_1)$, and one calculates the 
expansion of $(\det \cS)^{\star 2})$ into Schubert classes: 
\[ \begin{split} (\det \cS)^{\star 2})  \star \cO_{(2,1)}  = & -q t_2^2 t_4(t_1+t_4) \cO_\emptyset + q t_2 t_4 (t_2+t_3) (t_1+t_4) \cO_1\\ &
-q_{3d}  t_1 t_2 t_3 t_4 \cO_{(1,1)} -q_{3d}  t_1 t_2 t_3 t_4 \cO_{(2)} \\ &
+ (q_{3d} t_1 t_2 t_3 t_3 +  (t_2 t_4)^2) \cO_{(2,1)} - t_2 t_4^2(t_2 +t_3) \cO_{(2,2)} \/. \end{split} \]
Then 
\[  \chi( \cO_{(2,1)} \star (\det \cS)^{\star 2})= -t_2 t_3 t_4^2(1-q_{3d}) \/,\]
Then 
\[ \oint_{\Gr(2,4)} \cO_{(2,1)} =  -(t_1 t_2 t_3 t_4)^{-2} \cdot (t_2 t_3 t_4^2) = 
- (t_1^2 t_2 t_3)^{-1} \/, \]
as claimed in the conjecture. Similarly, one can check that 
\[   \chi( \cO_{(2,2)} \star (\det \cS)^{\star 2})= t_3^2 t_4^2(1-q_{3d}) \]
and again one may check directly the statement on the top form.
%The expansion into Schubert classes of $\det \mathcal{Q}$ is 
%\[
%\det \mathcal{Q} =
% t_3 t_4 \mathcal{O}_{\emptyset}+ t_2 t_4 \mathcal{O}_{(1)} + t_2 t_3 \mathcal{O}_{(2)} 
% + t_1 t_4 \mathcal{O}_{(1,1)} + t_1 t_3 \mathcal{O}_{(2,1)} + t_1 t_2 \mathcal{O}_{(2,2)}
%\]
%The expansion of $\det \mathcal{Q}^{\star 2}$ into Schubert classes is:
%\[
%\begin{split}
%\det \mathcal{Q}^{\star 2} = & t_4 t_3 \left( q^2 t_2 t_1 + (t_3 + t_4) (t_1 + t_2) q + t_3 t_4 \right) \mathcal{O}_\emptyset \\
%& + t_4 t_2 \left( (t_4 + (t_1 + t_3) q) t_2 + (q t_1 + t_3) t_4 + q t_1 t_3 \right) \mathcal{O}_{(1)} \\
%& + \left( (q t_2 + q t_3 + t_4) t_1 + (t_2 + t_3) t_4 + q t_2 t_3 \right) t_4 t_1 \mathcal{O}_{(1,1)} \\
%& + \left( (q t_1 + t_3 + t_4) t_2 + (q t_1 + t_4) t_3 + t_4 q t_1 \right) t_2 t_3 \mathcal{O}_{(2)} \\
%& + \left( (q t_2 + t_3 + t_4) t_1 + (t_4 + t_2) t_3 + t_2 t_4 \right) t_3 t_1 \mathcal{O}_{(2,1)} \\
%& + t_1 \left( (t_2 + t_3 + t_4) t_1 + (t_3 + t_4) t_2 + t_3 t_4 \right) t_2 \mathcal{O}_{(2,2)}
%\end{split}
%\]
%Collecting coefficients and multiplying by the nonzero top forms we obtain:
%\[ \begin{split} & (-t_2 t_3 t_4^2) \cdot \left( (q t_2 + t_3 + t_4) t_1 + (t_4 + t_2) t_3 + t_2 t_4 \right) t_3 t_1 \\
%& + (t_3 t_4)^2 \cdot  t_1 \left( (t_2 + t_3 + t_4) t_1 + (t_3 + t_4) t_2 + t_3 t_4 \right) t_2  \\
%& = (1-q) (t_1 t_2 t_3 t_4)^2 \end{split} \]
%Thus, 
%\[ \frac{ \oint_{\mathrm{Gr}(2,4)} \det \mathcal{Q}^{\star 2}}{(1-q)^2 (t_1 t_2 t_3 t_4)^2} = \frac{1}{1-q}  \/,\]
%as claimed.
\end{example}

\begin{example} We illustrate few examples in 
$\mathrm{Gr}(3,6)$. In this case, the top forms
$\oint_{\mathrm{Gr}(3,6)} \cO_\lambda$ are equal to $0$ 
unless $\lambda \supset (3,2,1)$, and otherwise:
\[ \begin{split} \oint_{\mathrm{Gr}(3,6)} \cO_{(3,2,1)} & =-(t_1^3 t_2^2 t_3^2  t_4 t_5)^{-1}\/; \quad
\oint_{\mathrm{Gr}(3,6)} \cO_{(3,2,2)} = (t_1^3 t_2^3 t_3 t_4 t_5)^{-1}\/; \\
\oint_{\mathrm{Gr}(3,6)} \cO_{(3,3,1)} & = (t_1^3 t_2^2 t_3^2 t_4^2)^{-1} \/; \quad
\oint_{\mathrm{Gr}(3,6)} \cO_{(3,3,2)}  = -\frac{t_3 + t_4}{t_1^3 t_2^3 t_3^2 t_4^2} \/; \\
\oint_{\mathrm{Gr}(3,6)}\cO_{(3,3,3)} & = (t_1^3 t_2^3 t_3^3)^{-1} \/. \end{split}
\]
The reader may observe that if one specializes $t_i \mapsto 1$ for $1 \le i \le 6$, then
one obtains the top form from \eqref{E:non-eq36}. The corresponding (quantum) Euler characteristics
$\chi_{\Gr(3,6)} (\cO_\lambda \star (\det \cS)^{\star 3})$ are equal to $0$ unless $\lambda \supset (3,2,1)$. The calculations of the latter are recorded below:
\[ \begin{split} \lambda = (3,2,1): &  -t_2 t_3 t_4^2 t_5^2 t_6^3(1 - q_{3d}) \/; \quad 
\lambda = (3,2,2): ~ t_3^2 t_4^2 t_5^2 t_6^3(1 - q_{3d}) \/; \\
\lambda = (3,3,1): & ~ t_2 t_3 t_4 t_5^3 t_6^3(1 - q_{3d}) \/; \quad 
\lambda = (3,3,2):  -t_3 t_4 t_5^3 t_6^3(t_3+t_4) (1 - q_{3d}) \/; \\
\lambda = (3,3,3): & ~ t_4^3 t_5^3 t_6^3(1 - q_{3d}) \/. \end{split} \]
%
% of $\det \mathcal{Q}^{\star 3}$ is significantly more complicated and will be omitted. A computer calculation shows that 
%\[ \frac{ \oint_{\mathrm{Gr}(3,6)} \det \mathcal{Q}^{\star 3}}{(1-q)^3 (t_1 t_2 t_3 t_4 t_5 t_6)^3} = \frac{1}{1-q} \]
%as claimed.
\end{example}

Finally, we mention that if one changes the Chern-Simons levels, the associated dictionary will be modified. Our claim is that the small (non)-equivariant quantum K theory of $\mathrm{Gr}(k;N)$ can be reconstructed from the quantum cohomology of the same target.  And it would also be interesting to verify our observation via integrable systems utilized in \cite{Korff:2010, Gorbounov:2014, Gu:2022ugf} and I/J-functions studied in \cite{Taipale:2013, Wen19, GY21, Yan21}.

\begin{center}
\section*{Acknowledgement}
\end{center}

We would like to thank Ming Zhang for useful discussions.  We would also like to thank Y.P. Lee for reading the draft and providing useful comments. Wei Gu was supported in part by NSF grant PHY-1720321; Jirui Guo was supported by the Fundamental Research Funds for the Central Universities; Leonardo Mihalcea was partially supported by NSF grant DMS-2152294 and a Simons Collaboration grant;  Yaoxiong Wen was supported by a KIAS Individual Grant (MG083902) at Korea Institute for Advanced Study and a POSCO Science fellowship; Xiaohan Yan was partially supported by ERC Consolidator Grant ROGW-864919 and by NSF grant DMS-1906326.
\\

\appendix
\section{The top form as a holomorphic Euler characteristic}\label{CGP}
%from classical case of Conjecturetop form A proof of a special case of equation (\ref{E:maintQH})}\label{CGP}
{In this section we verify the Conjecture \ref{conj:pairing} in the case $q=0$. The argument uses both a residue calculation and a 
Grothendieck-Riemann-Roch argument.} 
%In this section we verify \ref{E:maintQH} in the special case when $q=0$, utilizing the %Grothendieck-Riemann-Roch. 

For any $\kappa \in K(X)$ we have:
\begin{equation}\label{}\nonumber
  \chi (\kappa) =\int_{{\rm Gr}(k;N)}{\rm Ch}\left(\kappa \right){\rm Td}(T_{{\rm Gr}(k;N)}) \/,
\end{equation}
where $\mathrm{Ch}$ denotes the Chern character, {and $\mathrm{Ch}\left(\kappa \right)$ will be a polynomial of $e^{-L\sigma_{a}}$ that is denoted as ${\cal O}\left(e^{-L\sigma_{a}}\right)$}. One can use the Euler sequence for the Grassmannian to get the Todd class of the holomorphic tangent bundle of ${\rm Gr}(k;N)$ \cite{CKM14}:
 \begin{equation*}
   0\rightarrow {\cal S}\otimes{\cal S}^{\ast}\rightarrow\left({\cal S}^{\ast}\right)^{\oplus N}\rightarrow T_{{\rm Gr}(k;N)}\rightarrow 0.
 \end{equation*}
 Then one obtains:
 % find that
\begin{equation*}
  {\rm Td}(T_{{\rm Gr}(k;N)})=\prod^{k}_{a=1}\left(\frac{L\sigma_{a}}{1-e^{-L\sigma_{a}}}\right)^{N}\frac{\prod_{a\neq b}\left(1-e^{-L\left(\sigma_{a}-\sigma_{b}\right)}\right)}{\prod_{a\neq b}L\left(\sigma_{a}-\sigma_{b}\right)}.
\end{equation*}
Here $\sigma_a$'s are the Chern roots of $\mathcal{S}^*$.
We will calculate the integral above from the residue formula and the nonabelian/abelian formula
 for ${\rm Gr}(k;N)$ \cite{Martin:2000,BCK05}. For $g(\sigma) \in H^*(X)$,
%{\color{blue}Let us recall how to write integrals on the ${\rm Gr}(k;N)$ using resiude formula.} To get that, we first use the nonabelian/abelian correspondence in the intersection formula for ${\rm Gr}(k;N)$ \cite{Martin:2000,BCK05}
\begin{equation*}
  \int_{{\rm Gr}(k;N)} g\left(\sigma\right)=\frac{1}{k!}\int_{\left(\mathbb{P}^{N-1}\right)^{k}}\prod_{a\neq b}\left(\sigma_{a}-\sigma_{b}\right) g\left(\sigma\right).
\end{equation*}
Since there is a residue-type formula for calculating integrals:
%the intersection theory
on the projective space
\begin{equation*}
  \int_{\mathbb{P}^{N-1}}g(\sigma)=\oint_{\sigma=0}\frac{d\sigma}{2\pi i\sigma^{N}}g(\sigma).
\end{equation*}
 We then obtain that
\begin{equation}\label{E:ANA-app}
   \int_{{\rm Gr}(k;N)} g\left(\sigma\right)=\frac{1}{k!}\oint_{\sigma_a=0} \left(\prod^{k}_{a=1}\frac{d\sigma_{a}}{2\pi i \left(\sigma_{a}\right)^{N}}\right)\prod_{a\neq b}\left(\sigma_{a}-\sigma_{b}\right) g\left(\sigma\right).
\end{equation}
We now apply Grothedieck-Riemann-Roch to calculate:
%We now take
%{\color{blue} The above equation applies to a general function $g\left(\sigma\right)$. Thus, it is legitimate to make the following specification of $g\left(\sigma\right)$ in the above equation to compute the K-invariants}
%\begin{equation*}
%  g(\sigma)=\prod^{k}_{a=1}\left(\frac{L\sigma_{a}}{1-e^{-L\sigma_{a}}}\right)^{N}\frac{\prod_{b\neq c}\left(1-e^{-L\left(\sigma_{b}-\sigma_{c}\right)}\right)}{\prod_{b\neq c}L\left(\sigma_{b}-\sigma_{c}\right)} \cdot \mathrm{Ch}(a)
% % {\rm Ch}\left({\cal O}\right).
%\end{equation*}
%(so that we include the terms arising from the Grothedieck-Riemann-Roch) to obtain
%%{\color{red} LM: Remove $R$'s ? Also note I replaced $Ch(\mathcal{O})$ by an arbitrary $Ch(a) \in H^*(X)$. WG: I replace them by L}
%%So we have
\begin{eqnarray*}
\chi(\kappa)
   %\chi\left({\cal O}\right)
   &=& \int_{\Gr(k;N)} {\rm Td}(T_{\Gr(k;N)}) {\rm Ch}(\kappa) \\
   & = & \frac{1}{k!} \oint \left(\prod^{k}_{a=1}\frac{d\left(L\sigma_{a}\right)}{2\pi i \left(L\sigma_{a}\right)^{N}}\right)\prod_{b\neq c}\left(L\sigma_{b}-L\sigma_{c}\right){\rm Td}(T_{\Gr(k;N)}) {\cal O}\left(e^{-L\sigma_{a}}\right) \\
   &=&\frac{1}{k!} \oint_{\sigma_a=0} \left(\prod^{k}_{a=1}\frac{d\left(L\sigma_{a}\right)}{\left(1-e^{-L\sigma_{a}}\right)^{N}}\right)\prod_{b\neq c}\left(1-e^{-L\left(\sigma_{b}-\sigma_{c}\right)}\right){\cal O}\left(e^{-L\sigma_{a}}\right) \\
%   \left({\cal O}\right)  \\
   &=& \frac{1}{k!} \oint_{x_a=1} \left(\prod^{k}_{a=1}\frac{d(-x_{a})}{2\pi i \left(1-x_{a}\right)^{N}(x_{a})^{k}}\right)\prod_{b\neq c}\left(x_{b}-x_{c}\right){\cal O}\left(x_{a}\right) \\
%   \left({\cal O}\right) \\
   &=& \frac{1}{k!} \oint_{z_a=0} \left(\prod^{k}_{a=1}\frac{dz_{a}}{2\pi i (z_{a})^{N}(1-z_{a})^{k}}\right)\prod_{b\neq c}\left(z_{b}-z_{c}\right)
 {\cal O}\left(1-z_{a}\right)  \\
   &=& \int_{{\rm Gr}(k;N)}\frac{ {\cal O}\left(1-L\sigma_{a}\right) }
   %{\rm Ch}\left({\cal O}\right)}
   {\left(\prod^{k}_{a=1}1-L\sigma_{a}\right)^{k}}.
\end{eqnarray*}
In the fourth equality, we have changed the variable from $\sigma_a$ to $x_{a}=e^{-L\sigma_a}$, which induces a measure factor $\prod^{k}_{a=1}(x_{a})^{-1}$ in the integral. 
The last equality uses \eqref{E:ANA-app} above.

%{\color{red}In the last equality, we have changed the dumb variable from $z_a$ to $\sigma_a$ %such that the cohomological top-form integral is more explicitly.}
{Note that this calculation also explains the twisting
factor $\frac{1}{\left(\det S\right)^{k}}$. To relate to Conjecture \ref{conj:pairing}, 
we replace $\kappa$ by $\kappa \otimes (\det \mathcal{S})^k )$ to obtain: 
\[ \chi ( \kappa \otimes (\det \mathcal{S})^k ) = \chi ( \cO(e^{-L \sigma_a}) \otimes (\det \mathcal{S})^k )=  \int_{{\rm Gr}(k;N)} {\cal O}\left(1-L\sigma_{a}\right) \]
If $\kappa = \cO_\lambda \in K(\Gr(k;N))$ is a Schubert class represented by a Grothendieck polynomial
$G_\lambda(1-e^{-L \sigma_1}, \ldots, 1-e^{-L \sigma_k})$, then the right hand side of the previous expression becomes
\[ \int_{{\rm Gr}(k;N)} G_\lambda (L\sigma_1, \ldots , L \sigma_k) \/, \]
which is precisely the definition of the top form of $G_\lambda(\sigma)$. 
In other words, the tQH correlator of $\cO_\lambda$ is, on one side, the top form (as also verified in \eqref{E:topform}), and, on the other side, a holomorphic Euler characteristic. This verifies the statement in Conjecture \ref{conj:pairing} for $q=0$.}

%This finishes the proof of Eq.(\ref{E:maintQH}) in this special case, explaining along the way the twisting
%factor $\frac{1}{\left(\det S\right)^{k}}$.
%
%

\end{document}